\documentclass[twocolumn]{aastex62}
\usepackage[utf8]{inputenc}
\usepackage{amsmath}
\usepackage{graphicx}
\usepackage{lipsum}
\usepackage{lineno}
\usepackage[T1]{fontenc} % for Icelandic Characters
\usepackage[utf8]{inputenc} % for Icelandic Characters 
\newcommand{\teff}{\mbox{$\rm T_{\rm eff}$}}

\newcommand{\logg}{\mbox{$\log g$}}

\begin{document}

\title{Measuring the Temperature of Starspots from Multi-filter Photometry}
\author[0000-0003-2435-130X]{Maria C. Schutte}
\email{maria.schutte-1@ou.edu}
\affiliation{Homer L. Dodge Department of Physics and Astronomy, University of Oklahoma, 440 W. Brooks Street, Norman, OK 73019, USA}

\author{Leslie Hebb}
\affiliation{Department of Physics, Hobart and William Smith Colleges, 20 St. Clair Street, Geneva, NY 14456, USA}

\author[0000-0001-9209-1808]{John P. Wisniewski}
\affiliation{NASA Headquarters, 300 Hidden Figures Way SW, Washington, DC 20546, USA}

\author[0000-0003-4835-0619]{Caleb I. Ca\~nas}
\altaffiliation{NASA Postdoctoral Fellow}
\affiliation{NASA Goddard Space Flight Center, 8800 Greenbelt Road, Greenbelt, MD 20771, USA }

% \author[0000-0001-8401-4300]{Shubham Kanodia}
% \affiliation{Earth and Planets Laboratory, Carnegie Institution for Science, 5241 Broad Branch Road, NW, Washington, DC 20015, USA}

\author[0000-0002-2990-7613]{Jessica E. Libby-Roberts}
\affil{Department of Astronomy \& Astrophysics, 525 Davey Laboratory, The Pennsylvania State University, University Park, PA 16802, USA}
\affil{Center for Exoplanets and Habitable Worlds, 525 Davey Laboratory, The Pennsylvania State University, University Park, PA 16802, USA}

\author[0000-0002-9082-6337]{Andrea S.J.\ Lin}
\affil{Department of Astronomy \& Astrophysics, 525 Davey Laboratory, The Pennsylvania State University, University Park, PA, 16802, USA}
\affil{Center for Exoplanets and Habitable Worlds, 525 Davey Laboratory, The Pennsylvania State University, University Park, PA, 16802, USA}

\author{Paul Robertson}
\affiliation{Department of Physics and Astronomy, University of California - Irvine, 4129 Frederick Reines Hall, Irvine, CA 92697}

\author[0000-0001-7409-5688]{Guðmundur Stefánsson} 
\affil{NASA Sagan Fellow}
\affil{Department of Astrophysical Sciences, Princeton University, 4 Ivy Lane, Princeton, NJ 08540, USA}

\begin{abstract}
Using simultaneous multi-filter observations during the transit of an exoplanet around a K dwarf star, we determine the temperature of a starspot through modeling the radius and position with wavelength-dependent spot contrasts. We model the spot using the starspot modeling program STarSPot (\texttt{STSP}), which uses the transiting companion as a knife-edge probe of the stellar surface. The contrast of the spot, i.e. the ratio of the integrated flux of a darker spot region to the star’s photosphere, is calculated for a range of filters and spot temperatures. We demonstrate this technique using simulated data of HAT-P-11, a K dwarf (\teff = 4780 K) with well-modeled starspot properties for which we obtained simultaneous multi-filter transits using LCO’s MuSCAT3 instrument on the 2-meter telescope at Haleakala Observatory which allows for simultaneous, multi-filter, diffuser assisted high-precision photometry. We determine the average (i.e. a combination of penumbra and umbra) spot temperature for HAT-P-11's spot complexes is 4500 K $\pm$ 100 K using this technique. We also find for our set of filters that comparing the SDSS $g^{\prime}$ and $i^{\prime}$ filters maximizes the signal difference caused by a large spot in the transit. Thus, this technique allows for the determination of the average spot temperature using only one spot occultation in transit and can provide simultaneous information on the spot temperature and spot properties.
% Understanding magnetic activity on the surface of stars other than the Sun is important for both better understanding the magnetic activity of stars and is particularly important for planet hosting stars since exoplanet analyses must include these effects to properly characterize the exoplanet’s atmosphere. 
\end{abstract}

\section{Introduction}\label{sec:intro}

It is well accepted that the Sun’s myriad of activity signatures, including Sunspots, Sunspot cycles, flares, and coronal mass ejections are driven by Solar dynamo theory \citep{charbonneau2014}. Our proximity to the Sun enables us to quantify the complex short and long time-scale evolution of activity phenomena across a multitude of wavelength regimes, providing rich observational constraints on dynamo theory.
Sunspots form in groups and have a non-uniform temperature (e.g. \citealt{solanki2003}). The complex and varied sizes and shapes of Sunspot groups constrains the underlying magnetic activity driving their creation \citep{zirin1998}. 

Starspots, the equivalent of sunspots on the surfaces of other stars, are a cornerstone observable that constrains our understanding of magnetic activity levels, variations, and magnetic cycles of stars other than the Sun. Numerous techniques are capable of identifying and characterizing starspots, such as Doppler imaging \citep{vogt1987,barnes2001,barnes2015}, spectropolarimetry \citep{donati1997}, and long-term photometric and spectroscopic observations \citep{mcquillan2014,morris2017b,howard2021,anthony2022}. Starspots have been studied on a variety of stars, including giants \citep{berdyugina1998,oneal2004}, subgiants \citep{gosnell2022,schutte2022}, M dwarfs \citep{bergyugina2005,dave2015,robertson2020}, K dwarfs \citep{morris2017}, and young Solar analogues \citep{netto2020}. 

Spot sizes and spot temperatures are two key observational parameters that are useful for characterizing starspot activity. Quantifying these parameters has broad applicability to the interpretation of exoplanet transmission spectroscopy, whereby spot activity is known to contaminate efforts to characterize exoplanetary atmospheric properties \citep{alam2018,rackham2018,bruno2018,bruno2020}. Even small covering fractions of 1\%, starspots would be the largest source of contamination when trying to retrieve an exoplanet's atmosphere from transmission spectroscopy \citep{pont2007}. 

Spectroscopic starspot measurements on giant and subgiant stars are able to measure the darkest portion of the spot because the activity lines appear only at low temperatures, so their starspot temperature measurements are on the order of $\Delta T~=~1000~K$. Additionally, covering fractions (i.e. the total area across the surface of the star covered by spots) on giants and subgiants can be measured as upwards of $\sim$40\%. However, these methods lend themselves well to statistical studies and empirical relationships between the temperature of a star and the temperature of a starspot as shown in \citet{bergyugina2005} and \citet{herbst2021}. As noted in both papers, these empirical relations should be treated with caution as they fail to reproduce the well known solar spot temperature contrasts. While they should be applicable to all types of stars, it is best to be cautious especially around younger G type stars and older Solar analogues.

  Historically, quantifying the temperatures of starspots has been pursued using integrated optical spectroscopic observations of cool stars, for example by leveraging the different temperature sensitivities of TiO bands (i.e. TiO $\delta~8860\AA$) in the red-optical \citep{neff1995,oneal1996}. TiO only appears at much lower temperatures (e.g. $\Delta T \simeq 1000 K$) than the photosphere of G and K type stars \citep{oneal2004}, making it a particularly good diagnostic of spot temperatures for these stars. With the correct spot covering fraction and a model of an inactive M-dwarf spectrum for the spots and an inactive stellar temperature spectrum for the star, the active host star's spectrum can be reproduced. 
  
  It is more challenging to measure spot temperatures in M dwarfs using TiO lines as activity and the photospheric contributions blend \citep{bergyugina2005}. Spot temperature measurements typically require estimates to be made about the spot covering fractions at the time of spectroscopic observations, but starspot distributions can change dramatically over time, including over single stellar rotation periods, which can impact efforts to approximate covering fractions. Recently two-temperature spectral deconvolution analyses have constrained spot filling factors by the broad photometric modulations they produce \citep{gosnell2022} while using simultaneous spectra to measure the spot temperature.

High cadence, high precision photometric observations of starspot crossing events, whereby a companion transits a starspot or plage feature, enable precise measurements of the covering fraction, size, and number of starspots on the surface of the star \citep{wolter2009,morris2017, netto2020,schutte2022,3884}. Because there is a degeneracy between the sizes, location, and temperatures features measured via starspot crossing events, this method requires one to make an assumption of the temperature of the starspots. In this paper we explore use of high-precision, multi-filter photometric observations of starspot crossing events to simultaneously determine both temperatures and sizes of individual starspots.

HAT-P-11 is a K4 dwarf star with two planets, a close in Neptune-sized planet that orbits every 4.88 days \citep{bakos2010} and gas giant planet that orbits every $\sim~$9 years \citep{yee2018}. Only the Neptune-sized planet (HAT-P-11b) is known to transit the star, but the planet's orbital axis is oriented at an oblique angle compared to the star's spin axis. \citet{sanchis2011} found the sky-projected angle between the spin axis and orbital plane to be $\lambda~=~106\substack{+15 \\ -11}^{\circ}$ from the observations of the Rossiter-McLaughlin effect. %HAT-P-11 also has a measured rotation period of $\sim~$29 days which is similar to the Sun, but given it is a K dwarf star, it has a deeper convection zone, which leads to a more active star \citep{bergyugina2005}. In fact, the expected higher levels of activity are seen in the Kepler data as there are 
HAT-P-11 exhibits high level of activity, with 95\% of HAT-P-11b's \textit{Kepler} transits exhibiting starspot crossing events \citep{morris2017}. Modeling these starspot crossing events led \citet{morris2017} to determine that HAT-P-11's starspots were similar in physical size to Solar maximum sunspots but had covering fractions nearly two orders of magnitude higher than the Sun. %\citet{morris2017} found that HAT-P-11 has two active latitude bands similar to the Sun at Solar maximum. 
%Need to introduce STSP somewhere around here

One of the major assumptions used to model HAT-P-11's starspots was the spot contrast, i.e. the ratio of the integrated flux of the spot compared to the star's unspotted photosphere. %This contrast relies on the temperature of the spot, which was not known while \citet{morris2017} were modeling the starspots. 
\citet{morris2017} used the area-weighted contrast of sunspots (c = 0.3) which blends the sunspot umbra and penumbra temperatures with their appropriate areas. In this paper, we will describe our technique to simultaneously spot sizes and temperatures on HAT-P-11, using simultaneous multi-band transit photometry obtained with LCOGT's MuSCAT3 instrument on the 2.0-meter telescope at Haleakala Observatory.

\section{Methods and Analysis}\label{sec:params}

\subsection{Modeling Active Regions on HAT-P-11}\label{sec:STSP}

\citet{morris2017} used the program STarSPot\footnote{The code for STarSPot can be found here: https://github.com/lesliehebb/STSP.} (\texttt{STSP}) to model the starspot crossing events in \textit{Kepler} observations of HAT-P-11. \texttt{STSP}'s functionality has been described and applied to model starspot characteristics in other cool star systems \citep{dave2015, morris2017,wisniewski2019,schutte2022,3884}. \texttt{STSP} generates synthetic light curves for a star using a pre-defined number of static, non-overlapping spots or spot complexes, and computes spot properties from a $\chi^{2}$ comparison between observed and synthetic fluxes using an affine-invariant Markov Chain Monte Carlo (MCMC) based on \citet{foreman2013}.

\subsection{Synthetic Spectra and Theoretical Contrast}

\begin{figure*}[t]
%  \begin{subfigure}[t]{.4\textwidth}
\centering
\includegraphics[width=.45\linewidth]{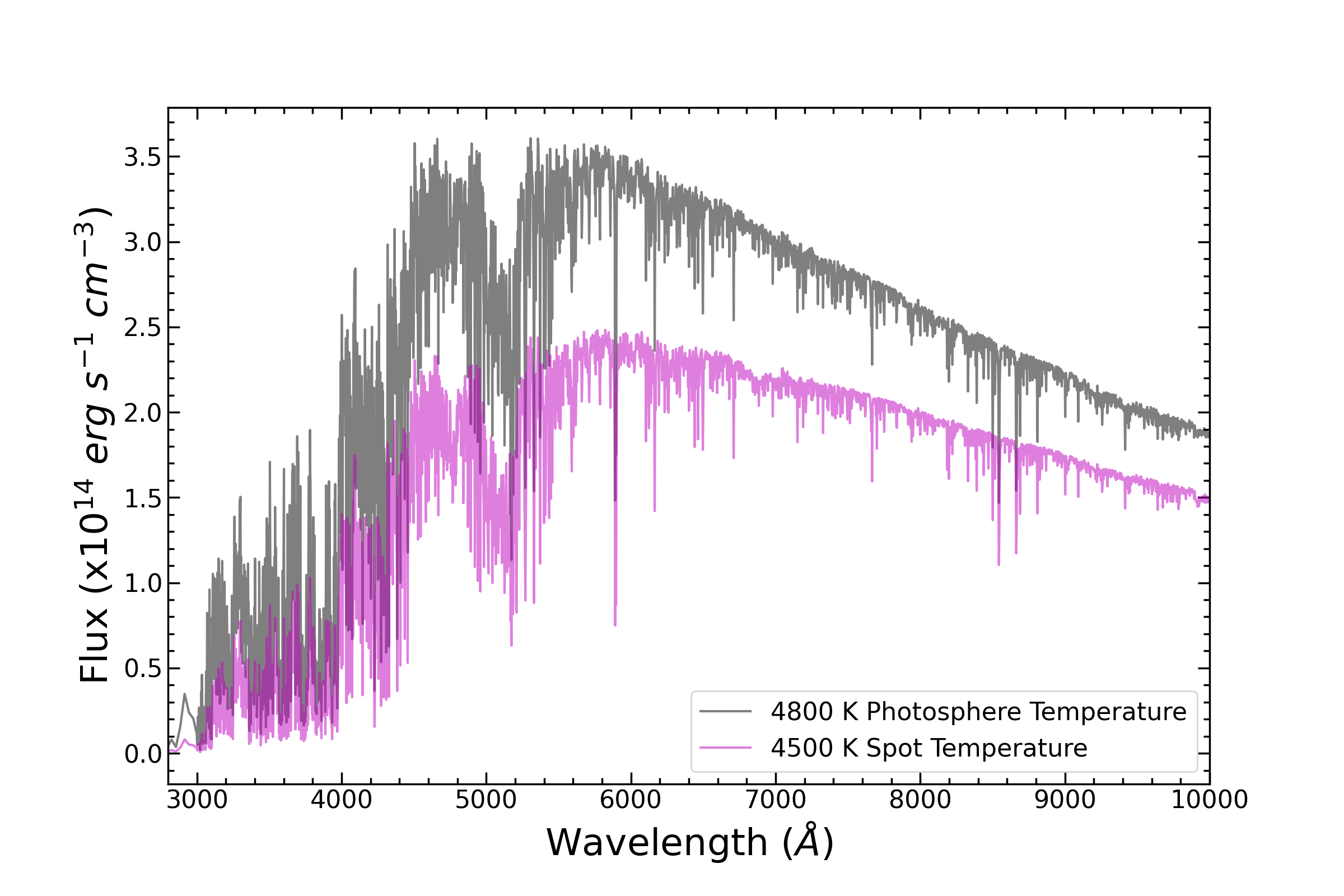}
\includegraphics[width=.45\linewidth]{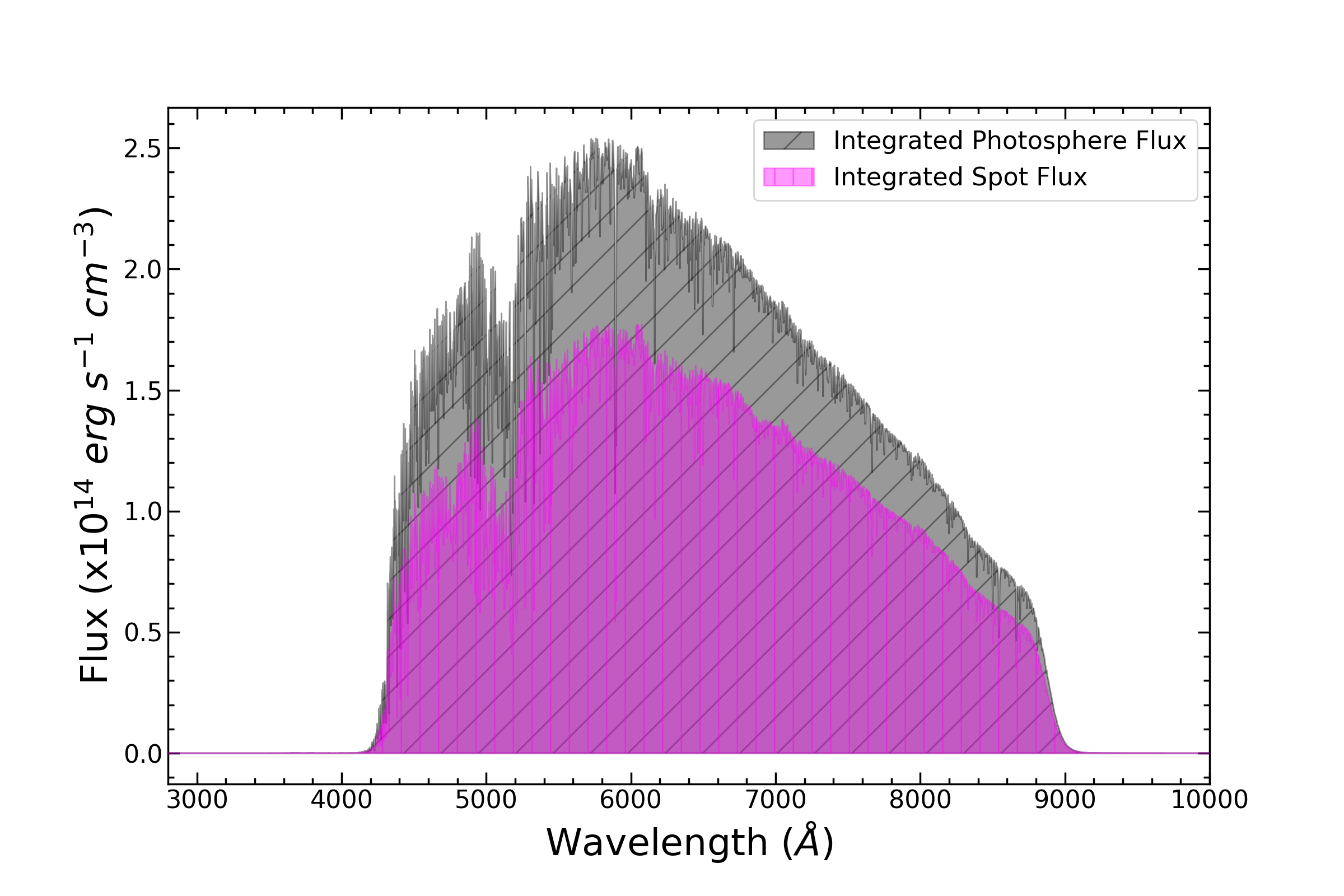}
\caption{{\bf Left:} HAT-P-11's stellar photosphere spectrum with $\teff = 4800~K$ and $\logg = 4.5$ in grey. Spectrum for spot temperature of 4500 K and $\logg = 4.5$ is shown in magenta. Spectra were obtained using the \texttt{expecto} python package, which provides a PHOENIX model spectrum for the closest grid point for an input effective temperature and surface gravity for solar metallicity stars. {\bf Right:} The fluxes over which the spectra were integrated for the \textit{Kepler} filter for the photosphere and spot are shown in filled in hatches of grey and magenta respectively. Once integrated over, the hatched regions correspond to the $I_{spot}$ (magenta) and $I_{phot}$ (grey) in Equation 1 which becomes a contrast of 0.32.}
\label{fig:spectrum}
\end{figure*}

The contrast of a starspot is approximated as: 
\begin{equation}
c = 1 - I_{spot}/I_{phot}
\end{equation}
where $I_{spot}$ is the integrated flux of the spot in a given wavelength range and $I_{phot}$ is the integrated flux of the photosphere in the same wavelength range \citep{solanki2003}.
Previously when modeling the active regions of HAT-P-11, \citet{morris2017} adopted the area weighted mean sunspot contrast of 0.3, as this contrast was found to best fit the data when they tested a range of contrasts ($c =$ 0.15-0.8). The area-weighted contrast takes into account the difference in contrast and area between a sunspot's umbral and penumbral regions with the umbra being much darker but also much smaller. \citet{morris2017} used a mean umbral contrast of 0.65 and mean penumbral contrast of 0.2 with the penumbra having an area around four times larger than the umbra \citep{solanki2003}, which provides an area-weighted sunspot contrast of 0.3. Then, after testing the contrast of 0.3 along with contrasts of 0.15, 0.6 and 0.8, \citet{morris2017} found that 95\% of spots in the data were fit with the area-weighted sunspot contrast.

In order to theoretically determine the spot contrast needed for a filter, a synthetic spectrum is needed for both the photosphere's temperature and for the spot's temperature, with the surface gravity assumed to be the same across the entire region of the star. We used the python package \texttt{expecto} which retrieves PHOENIX \citep{phoenix} synthetic model stellar spectra for the closest grid point to the input photosphere's temperature and surface gravity (i.e., the effective temperature is rounded to the closest 100 K temperature and the surface gravity is rounded to the nearest half), though it only allows for solar metallicity spectra.\footnote{https://expecto.readthedocs.io/en/latest/index.html}. Figure \ref{fig:spectrum} shows the PHOENIX synthetic spectrum for HAT-P-11's stellar photosphere. 

After obtaining synthetic spectra for the photospheric temperature and a range of spot temperatures, we multiplied each spectrum with the filter response curve for each bandpass of interest. The filter response curves were obtained using the SVO filter profile service \citep{rodrigo2012,rodrigo2020}. For our purposes, we used the following filters: SDSS $g^{\prime}$, $r^{\prime}$, $i^{\prime}$, and $z^{\prime}$ \citep{sdss};  \textit{Kepler} \citep{kepler}; TESS \citep{tess}; and OAO Zs \citep{muscat3}. Once we multiplied each spectra for all of the filters, we finally integrated over the wavelength region of the filter for both the stellar photosphere and the starspot temperature.

\begin{figure*}[t]
\centering
\includegraphics[width=0.95\textwidth]{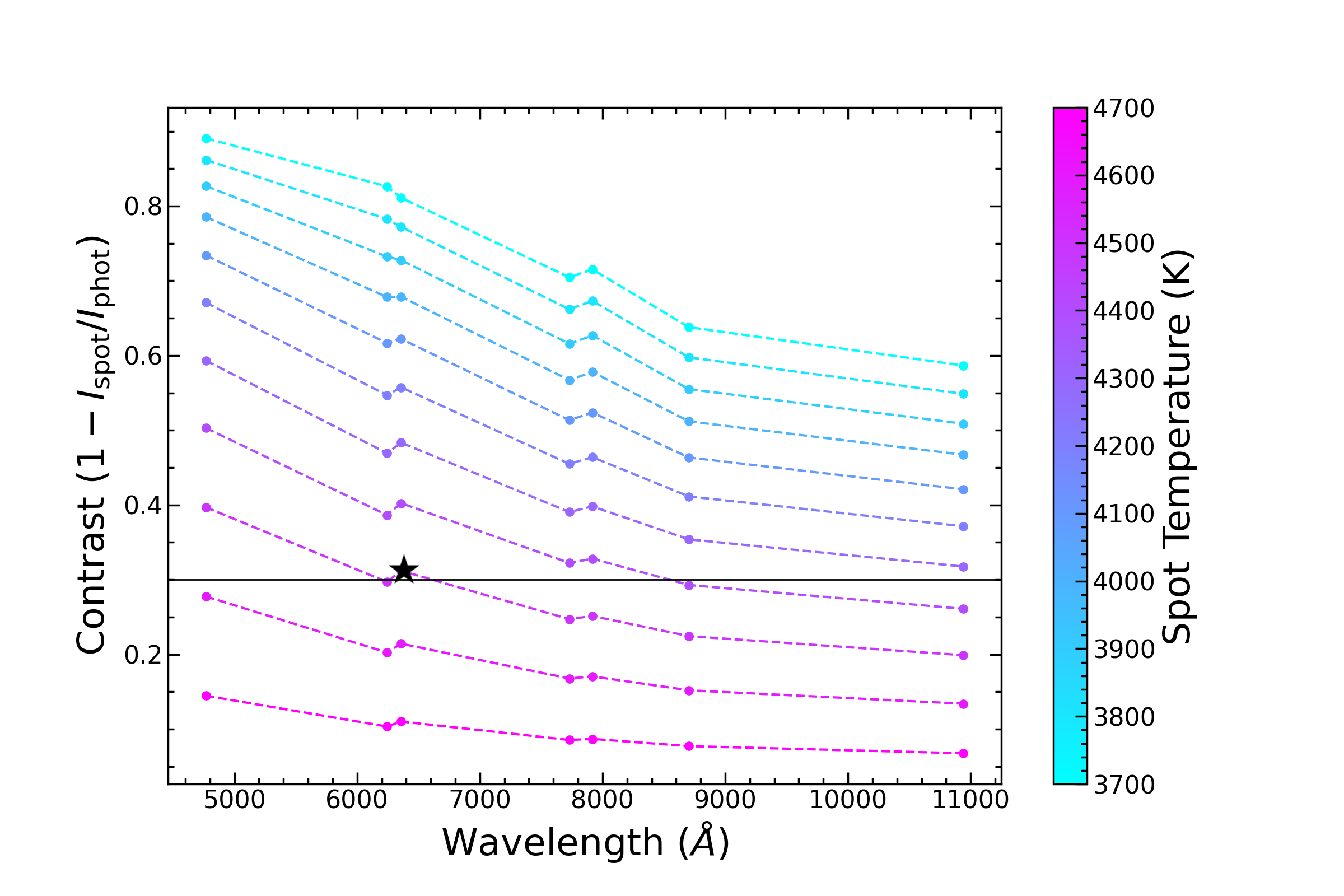}
\caption{Theoretical contrast values for HAT-P-11 assuming $\teff = 4800~K$ and $\logg = 4.5$. A range of spot temperatures from 3700-4700 K in 100 K steps were used to calculate the contrast of the starspot for a range of filters. The contrast values are plotted as points at the filter's central wavelength, though the contrast applies to the entire bandpass. In order of central wavelength, the filters used were: SDSS $g^{\prime}$, SDSS $r^{\prime}$, \textit{Kepler}, SDSS $i^{\prime}$, TESS, OAO Zs, and SDSS $z^{\prime}$. The solid black line at 0.3 corresponds to the area-weighted sunspot contrast assumed in \citet{morris2017} for their starspot modeling. The lines are colored by their spot temperature with the coolest spots in cyan. The black star corresponds to the contrast value for magenta spot spectrum shown in Figure \ref{fig:spectrum}.}
\label{fig:con_curves}
\end{figure*}

%What to cite for semrock filter? Also need to add in other citations for filters and SVO
For HAT-P-11, we used a stellar photosphere temperature of 4780 K (rounded to 4800 K) and a surface gravity of 4.59 \citep{bakos2010} (rounded to 4.5). We used a range of active region temperatures starting at 3700 K and increasing by 100 K until we reached 4700 K. The calculated contrasts for each spot temperature and filter are shown in Figure \ref{fig:con_curves} with contrast values plotted at their central wavelength. When looking at the \textit{Kepler} filter, the closest spot temperature that matches the data is 4500 K (black star indicates this point), which corresponds to a contrast value of 0.32. The metallicity of a star will also affect the theoretical stellar spectrum, so we explored using a more metal rich theoretical stellar spectrum to better match HAT-P-11's metallicity of [Fe/H] = +0.31. Using SVO's theoretical stellar spectra service, we obtained a BT-Settl synthetic spectrum for HAT-P-11's metallicity, surface gravity, and all photosphere and spot temperatures. With the different metallicity spectra, we performed the exact same procedure as just described and found the contrast values themselves are changed on the order of 0.01\%, so for HAT-P-11, the metallicity does not affect the contrast values significantly, though this could impact other stars of even higher or lower metallicity more substantially.

\section{LCO MuSCAT3 Observations}\label{sec:data}

\subsection{MuSCAT3 Simultaneous Multi-filter Photometry}

Using the theoretical contrast values for a variety of spot temperatures for HAT-P-11, the next step is to compare those contrast values using data from multiple filters including the \textit{Kepler} bandpass \citet{morris2017} used in modeling the \textit{Kepler} transits. However, to ensure that we are comparing the same active region occultation in each filter, simultaneous multi-filter photometry is needed. One such example of this type of instrument is LCO's MuSCAT3 instrument on the 2.0-meter FTN telescope at Haleakala Observatory \citep{muscat3}. MuSCAT3 allows for simultaneous multi-band photometry in four filters: SDSS $g^{\prime}$, $r^{\prime}$, and $i^{\prime}$ and OAO Zs. Unique to MuSCAT3 are the available engineered diffusers for each of the four filters. These types of diffusers increase the precision of ground-based photometry of transiting systems to the level of space-based data \citep{stefansson2017}. For a bright target like HAT-P-11, the diffusers allow for the precision needed to potentially observe a starspot occultation. We were awarded time with this instrument to observe two nights to increase our chances of observing a starspot crossing event. We observed one full transit of HAT-P-11b on June 29th, 2021.
%one full transit of Kepler-63 on 05-25-2021 (see Appendix for more information on this transit)

\subsubsection{HAT-P-11}

\begin{figure}[h]
\centering
\includegraphics[width=0.48\textwidth]{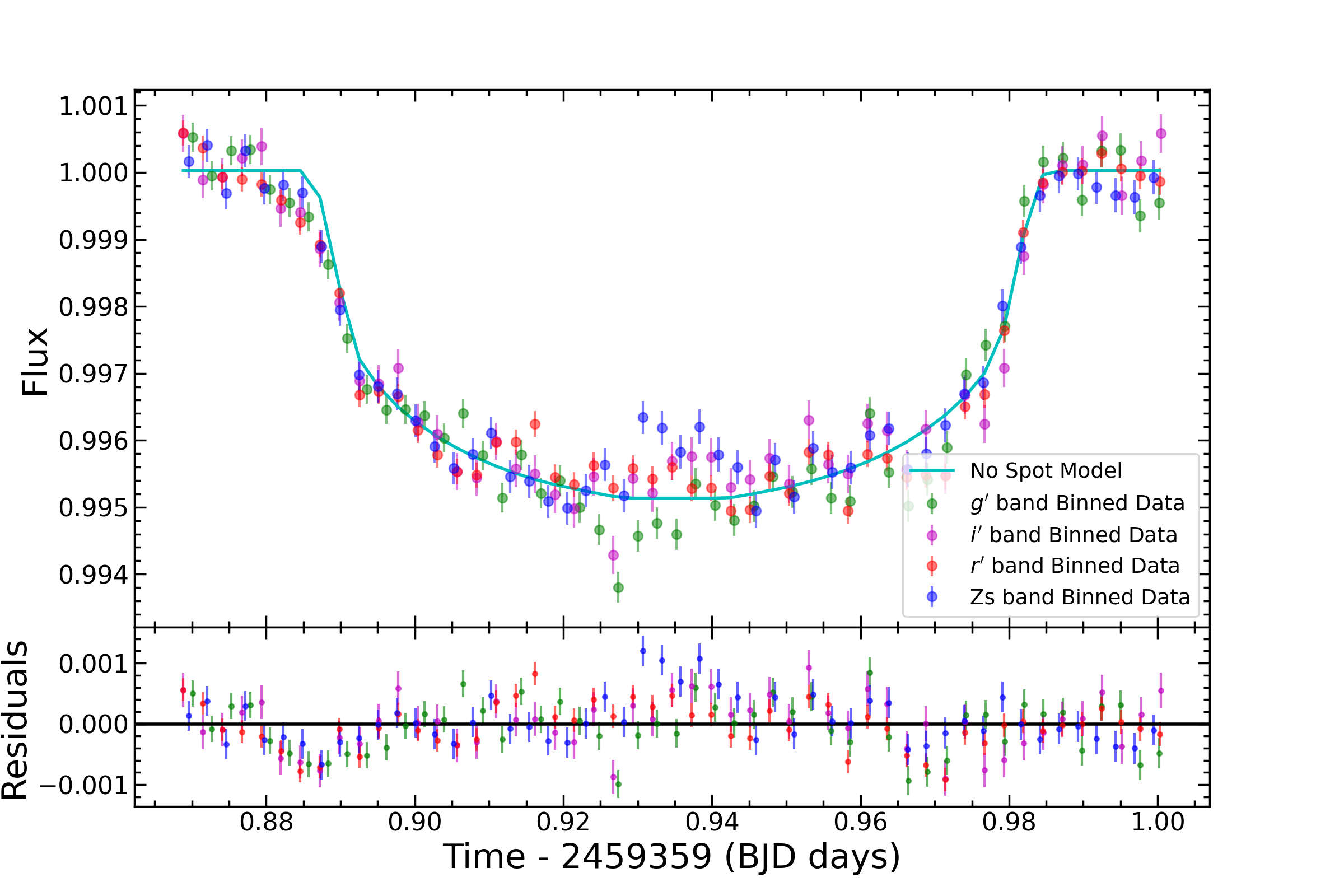}
\caption{Data from LCO MuSCAT3 from HAT-P-11b transit on June 29th, 2021 binned to 200 seconds for SDSS $r^{\prime}$ (red) and SDSS $i^{\prime}$ (magenta) and 210 seconds for SDSS $g^{\prime}$ (green) and Zs (blue) with all filters plotted on top of each other. The bottom panel shows the residuals of the no spot model for that filter minus the binned data points (cyan line here shows example no spot model for SDSS $r^{\prime}$). }
\label{fig:all_bin}
\end{figure}

On June 29th, 2021, we obtained a full transit of HAT-P-11b using the MuSCAT3 instrument. With magnitudes around 9 in all four filters (SDSS $g^{\prime}$, $r^{\prime}$, and $i^{\prime}$ and OAO Zs), we opted to use available diffusers for each filter as the diffusers reduce the scintillation noise. Due to the nature of the instrument, the light from the star is split into four, so the amount of light in each filter is reduced compared to a single filter CCD. Because of this fact, the exposure times used were 40 seconds in the $r^{\prime}$ and $i^{\prime}$ filters and 70 seconds in the $g^{\prime}$ and Zs filters. These long exposures lead to peak counts of 65,000-100,000 ADU, which is well below the saturation limit for each CCD ($>$ 100,000 ADU). These data were automatically processed using the LCO BANZAI pipeline \citep{muscat3}. We then ran each individual filter through the multi-aperture photometry process using AstroImageJ \citep{collins2017}. The precision for this transit was excellent with all filters having $<$1 mmag precision. There is one section of all four light curves that is slightly worse at the very start of the observation, which is attributed to partly cloudy weather at that time. Even after removing the poor section at the beginning of the night, it does not appear that there was a visible active region occultation at any point in the transit.

In order to further constrain the limits on a spot crossing event in the transit, we binned the data to 200 seconds for the SDSS $r^{\prime}$ and $i^{\prime}$ filters and 210 seconds for the SDSS $g^{\prime}$ and Zs filters and plotted all of the filters at once in Figure \ref{fig:all_bin}. We also plotted the residuals of the the binned data with respect to the no spot transit model for each filter with each filter having different four-parameter limb darkening coefficients but otherwise consistent transit model parameters \citep{claret}. For Figure \ref{fig:all_bin}, we chose to show one example no spot transit model in cyan for the SDSS $r^{\prime}$ transit. From the residuals, we see no evidence of a color-dependent spot crossing event during this transit, though there is a bump around mid-transit present only in the Zs filter which is further discussed in Section \ref{sec:simul}. Even with the excellent precision data (less than 1 mmag in every filter's unbinned data), there is a possibility that there was a spot in the path of the planet during this transit, but the spot was too small to detect.

\subsection{Simulated Light Curves}\label{sec:simul}

\begin{figure}[h]
    \centering
    \begin{minipage}{0.35\textwidth}
        \centering
        \includegraphics[width=\textwidth]{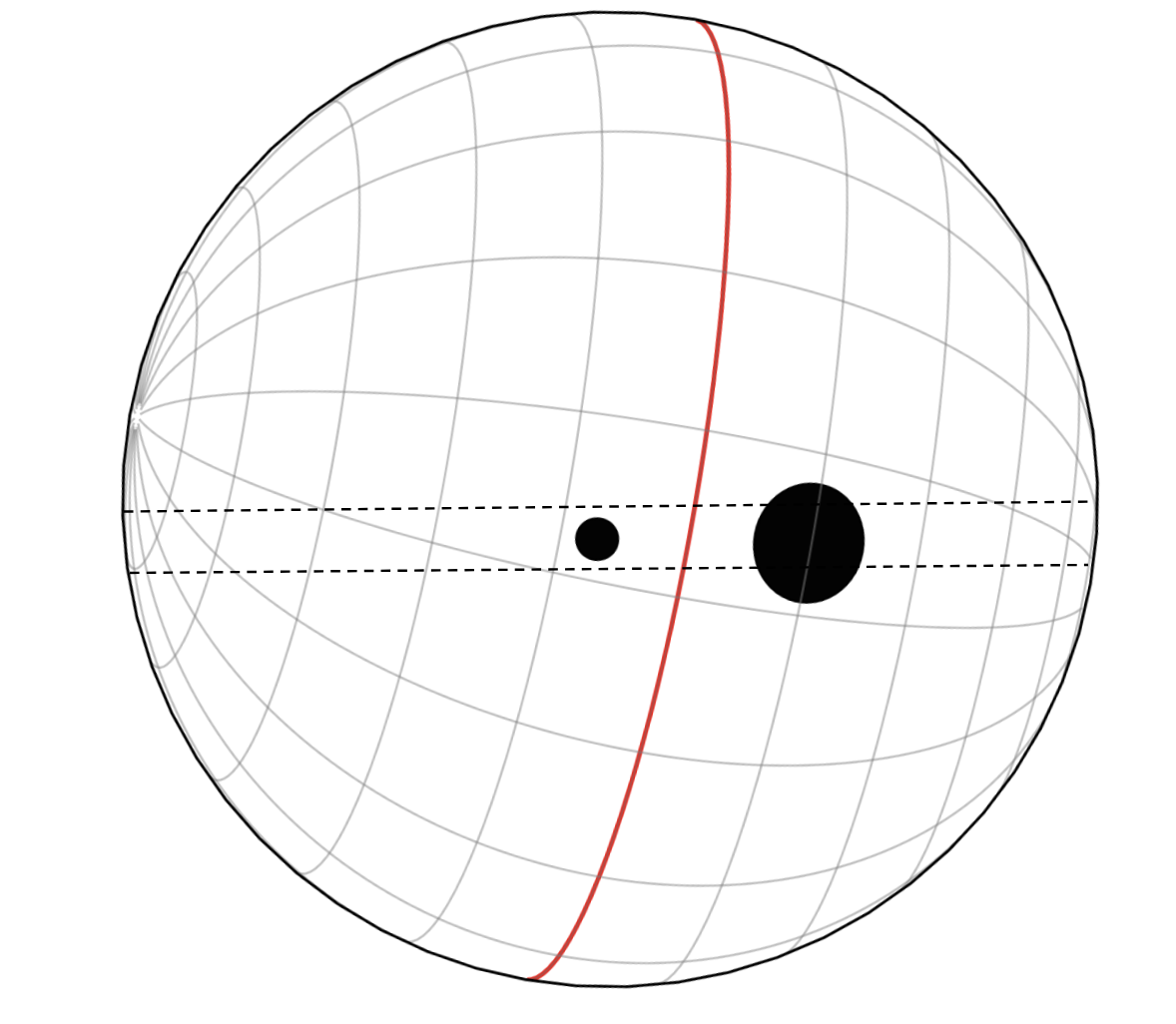} 
    \end{minipage}%
    \begin{minipage}{0.48\textwidth}
        \centering
        \includegraphics[width=\textwidth]{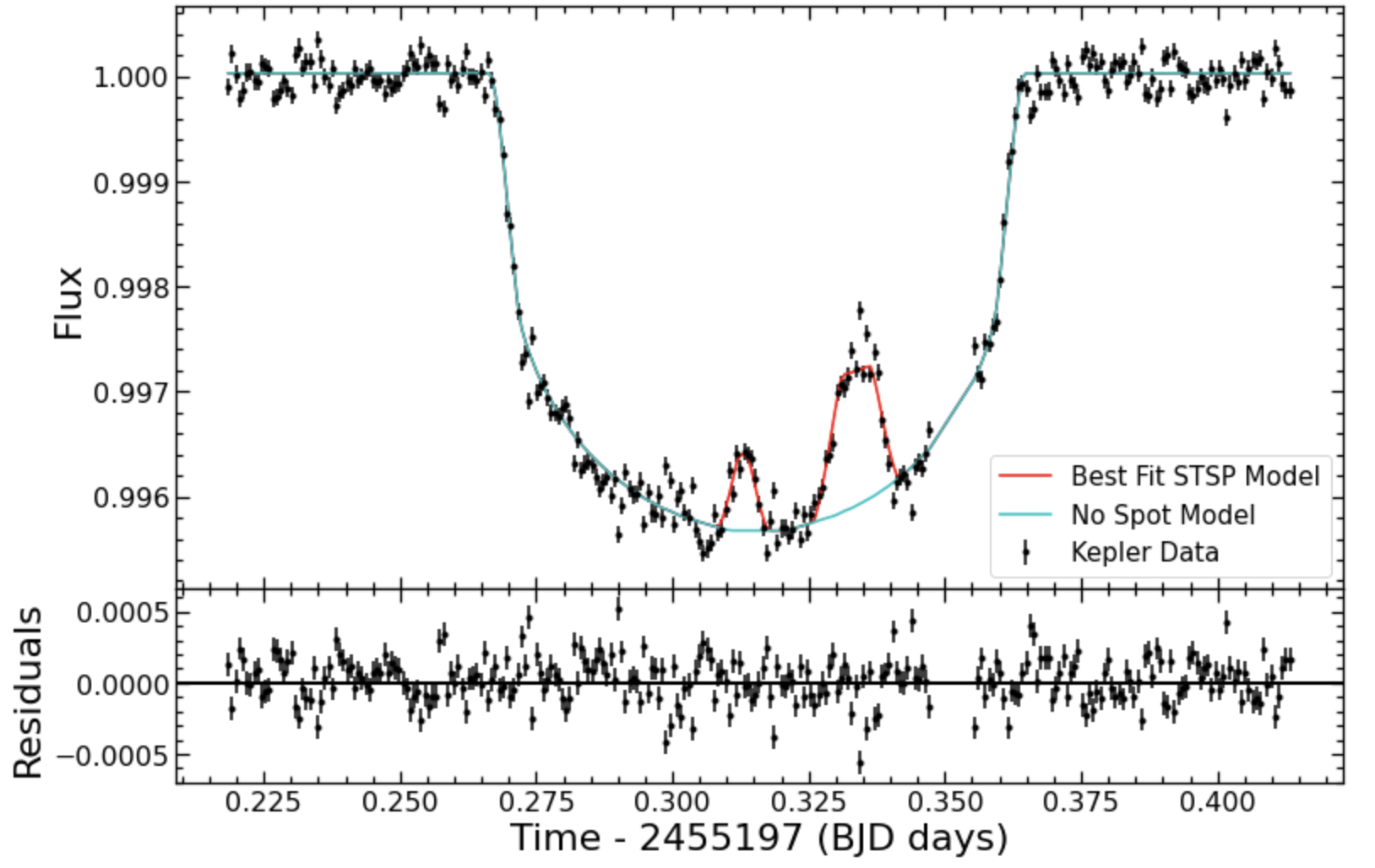} 
        \caption{{\bf Top:} Plot of the surface of HAT-P-11 with the final spot groups shown as black filled circles along with the red line denoting the equator of the star and with dashed lines denoting the full extent of the transit path for the secondary object (given $\lambda~=~106\substack{+15 \\ -11}^{\circ}$). {\bf Bottom:} Light curve for final \texttt{STSP} fit (red line) along with the no spot model for HAT-P-11 (cyan line) for chosen transit. The residuals (model - data) are shown below the light curve with black points.}\label{fig:hat11_transit}
    \end{minipage}
\end{figure}

Since we did not observe a starspot crossing occultation in our MuSCAT3 transit of HAT-P-11b, we instead looked to \textit{Kepler} transits of the planet to test our theoretical contrast values for various filters. First, we chose a transit of HAT-P-11b that had one unusually large starspot and one smaller, more typical size starspot for HAT-P-11 (see Figure \ref{fig:hat11_transit}). Considering the starspots for this transit have already been modeled by \citet{morris2017} using \texttt{STSP}, we know the radii and locations for the two spots as shown in Figure \ref{fig:hat11_transit}. Using this information and our theoretical contrast values, we can model how the spot occultation changes with different contrast values using \texttt{STSP}. We can then use this data to estimate what we should have been able to observe with MuSCAT3 data if there were similar size spots in the transit we obtained. 

\begin{figure}[h]
\centering
\includegraphics[width=0.48\textwidth]{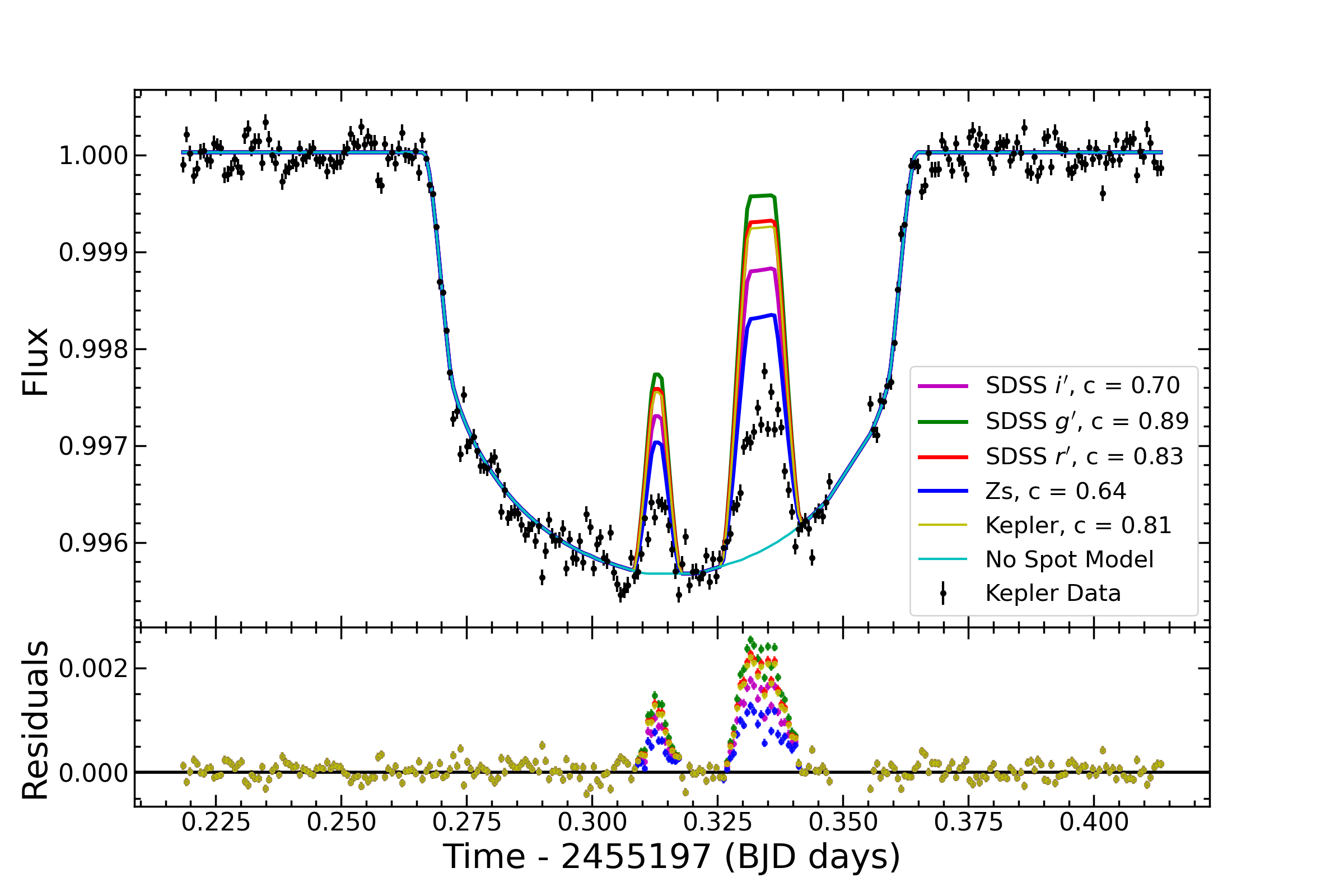}
\caption{Starspot models using theoretical contrast curves assuming spot temperature of 3700 K for SDSS $g^{\prime}$ (green line), SDSS $r^{\prime}$ (red line), SDSS $i^{\prime}$ (magenta line), and Zs (blue line) filters compared to \textit{Kepler} data (black points) and the no spot transit model (cyan line). This spot temperature assumes the spot temperature difference versus photosphere temperature model from \citet{herbst2021}. This spot temperature produces too dark spots (bumps are too big) to fit the data.}
\label{fig:kep_3700}
\end{figure}

\begin{figure}[h]
\centering
\includegraphics[width=0.48\textwidth]{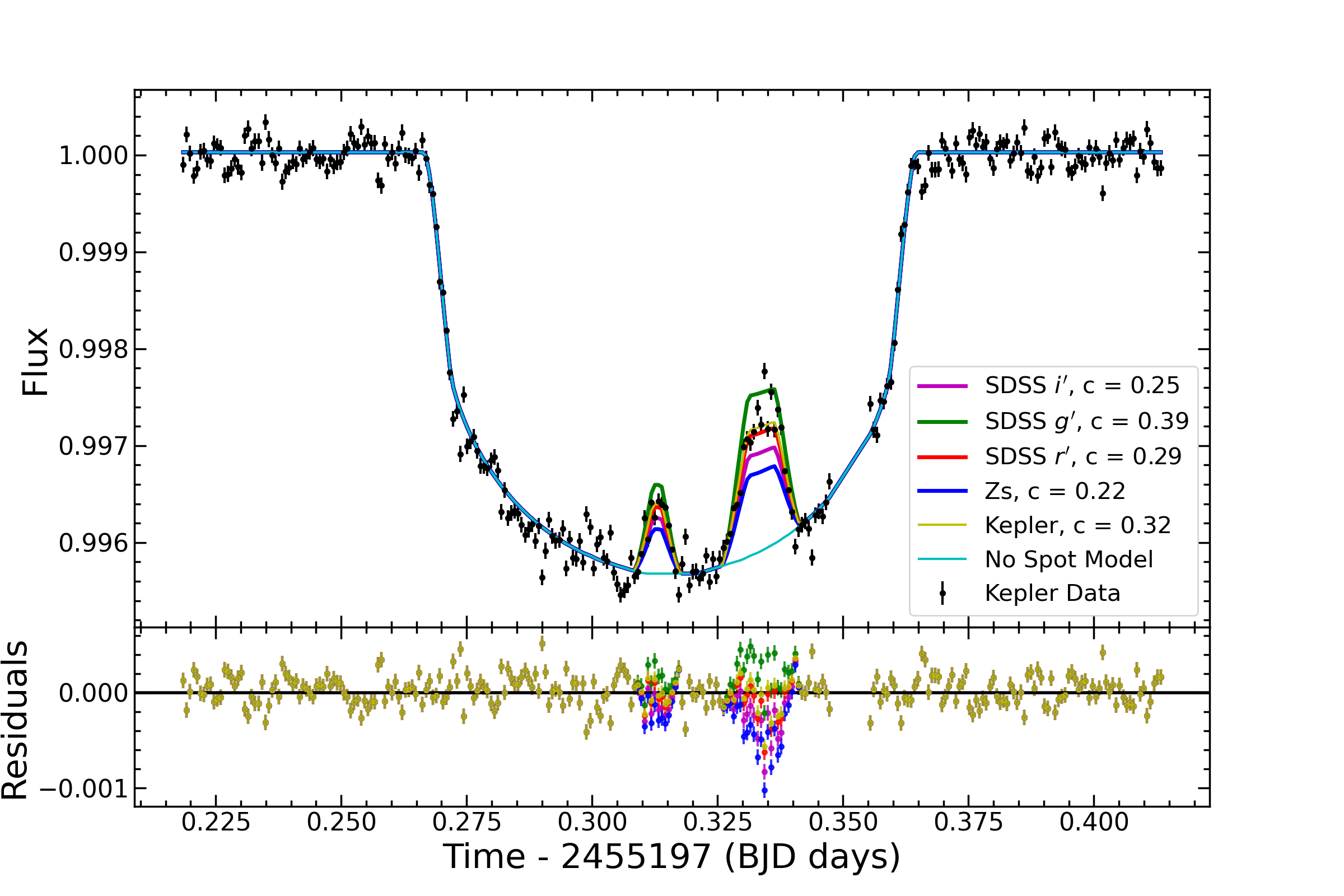}
\caption{Starspot models using theoretical contrast curves assuming spot temperature of 4500 K for SDSS $g^{\prime}$ (green line), SDSS $r^{\prime}$ (red line), SDSS $i^{\prime}$ (magenta line), \textit{Kepler} (yellow line) and Zs (blue line) filters compared to \textit{Kepler} data (black points) and the no spot transit model (cyan line). This spot temperature fits the \textit{Kepler} data best and corresponds to a contrast of 0.32 in the \textit{Kepler} band, similar to the value of 0.3 assumed in \citet{morris2017}.}
\label{fig:kep_4500}
\end{figure}

First, we must assume a spot temperature in order to know which contrast values to use for the four MuSCAT3 filters. If we were to use the empirical equations from \citet{herbst2021} to estimate the temperature of the active regions of HAT-P-11, the spot temperature would be around 3700-3800 K (a difference of $\sim~$1000 K from the photosphere). However, as seen in Figure \ref{fig:kep_3700}, the \texttt{STSP} models for all of the filters including the \textit{Kepler} bandpass result in starspot crossing events that are too high and do not match the data well as these methods determine the temperature for the darkest starspot regions rather than the average temperature. If we instead find the spot temperature that best matches the \textit{Kepler} data, the spot temperature would be 4500 K as seen in Figure \ref{fig:kep_4500}, which also shows that the \texttt{STSP} models are much closer to the correct height. As a spot temperature of 4500 K matches the data the best, we will use that spot temperature for the rest of the paper. Though a 4500 K spot temperature best matches the data, there is no way to confirm with only one filter, and if the spot radii and positions changed slightly, the best fit spot temperature could change.

\begin{figure*}[t]
%  \begin{subfigure}[t]{.4\textwidth}
\centering
\includegraphics[width=.4\linewidth]{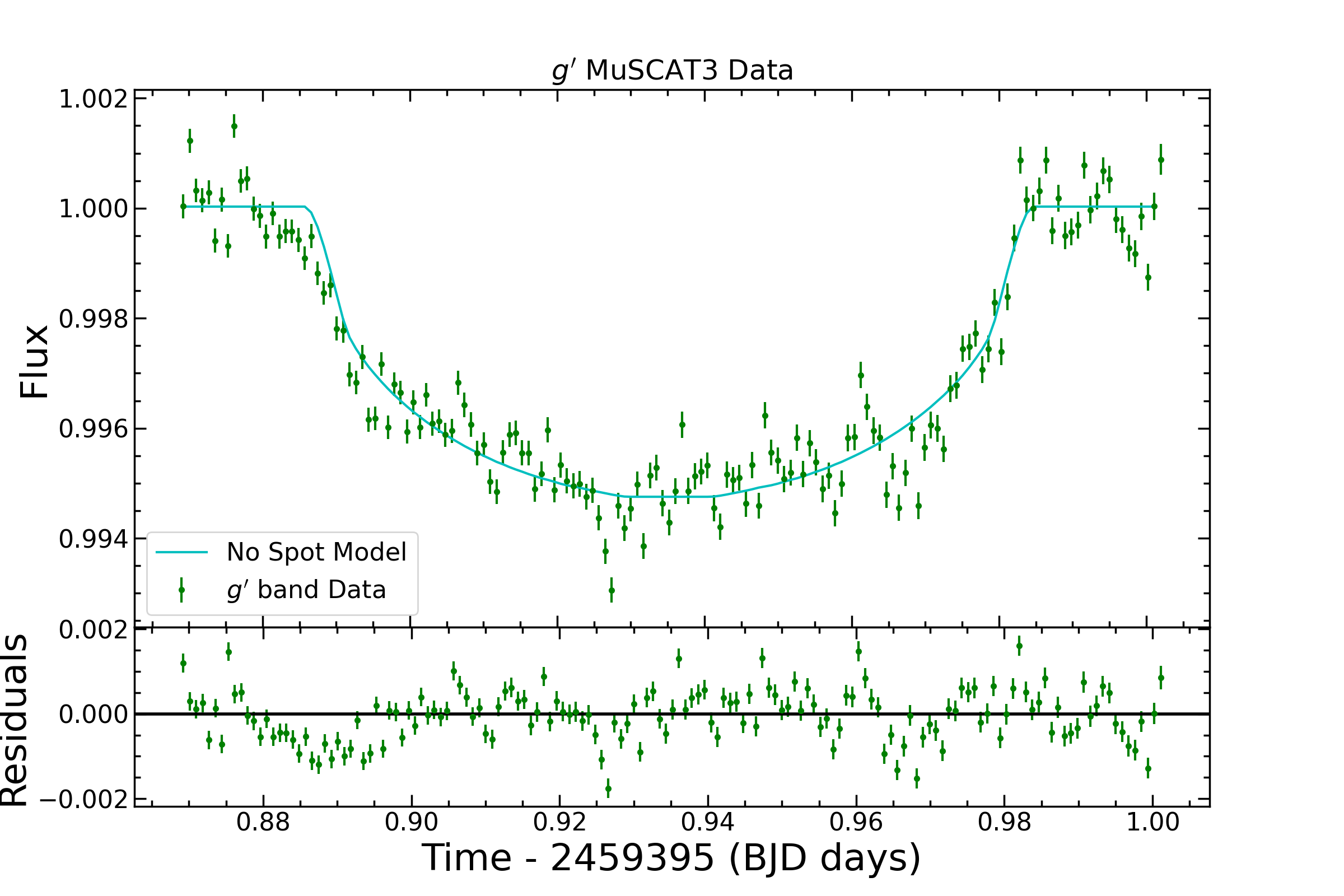}
\includegraphics[width=.4\linewidth]{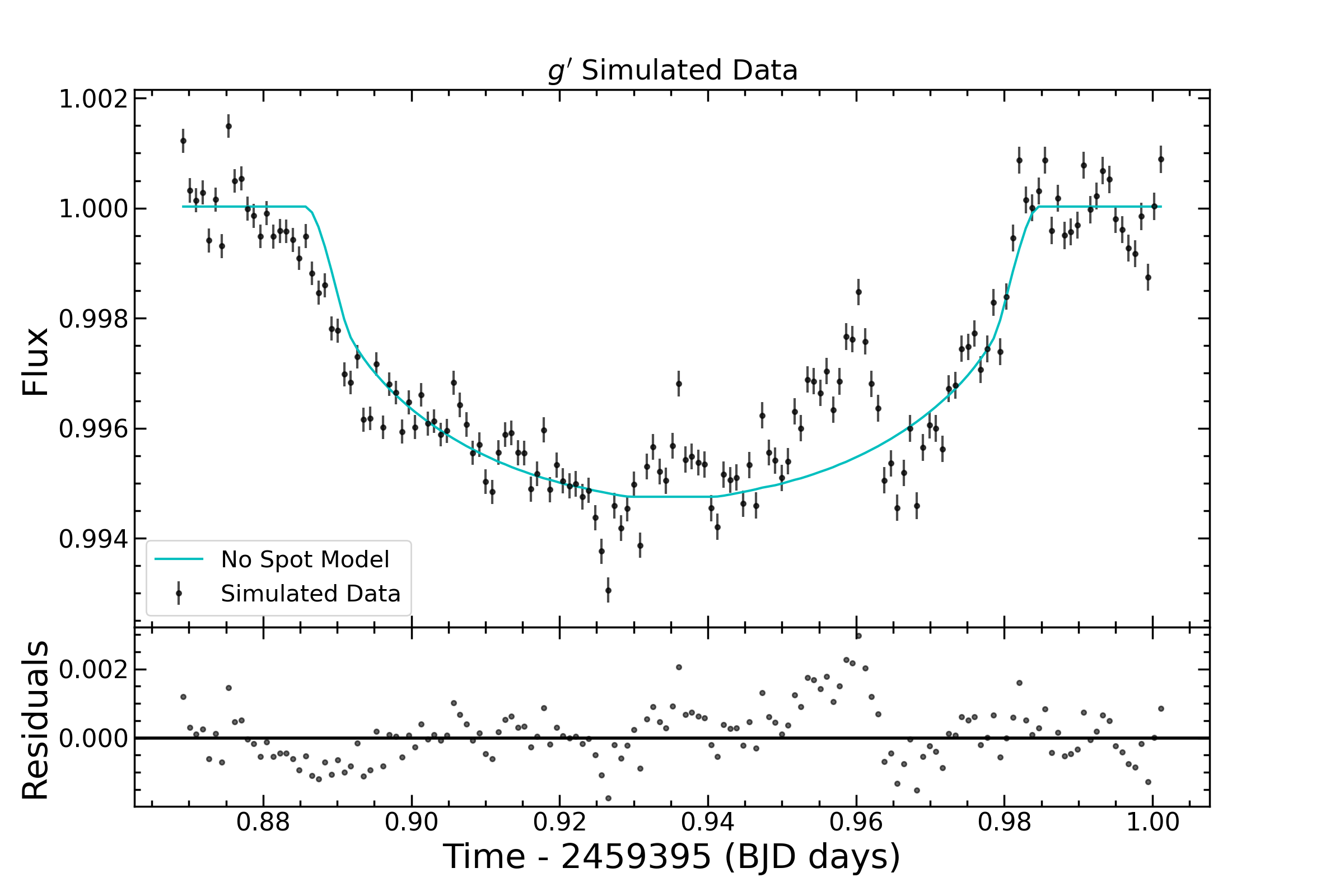}
\includegraphics[width=.4\linewidth]{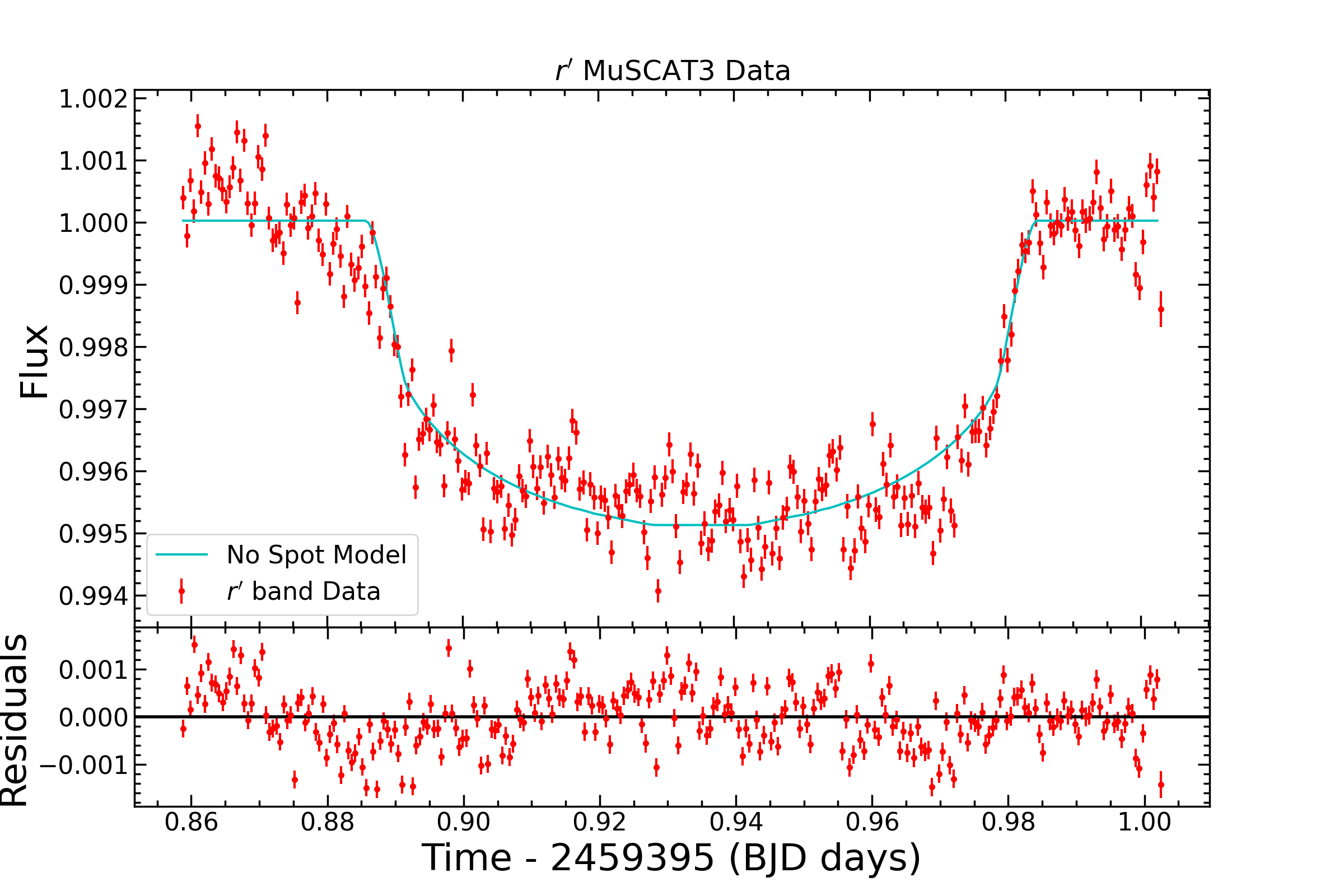}
\includegraphics[width=.4\linewidth]{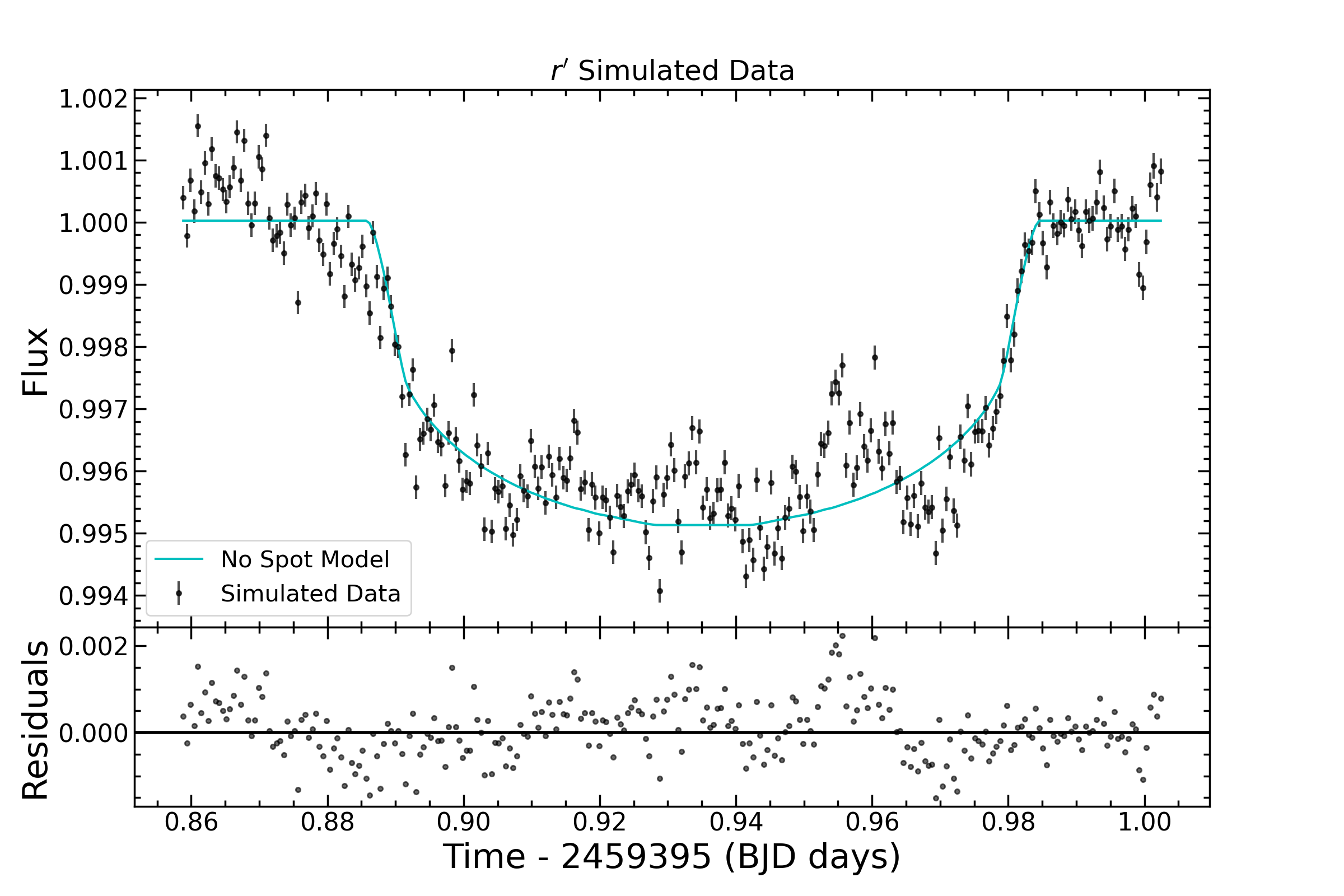}
\includegraphics[width=.4\linewidth]{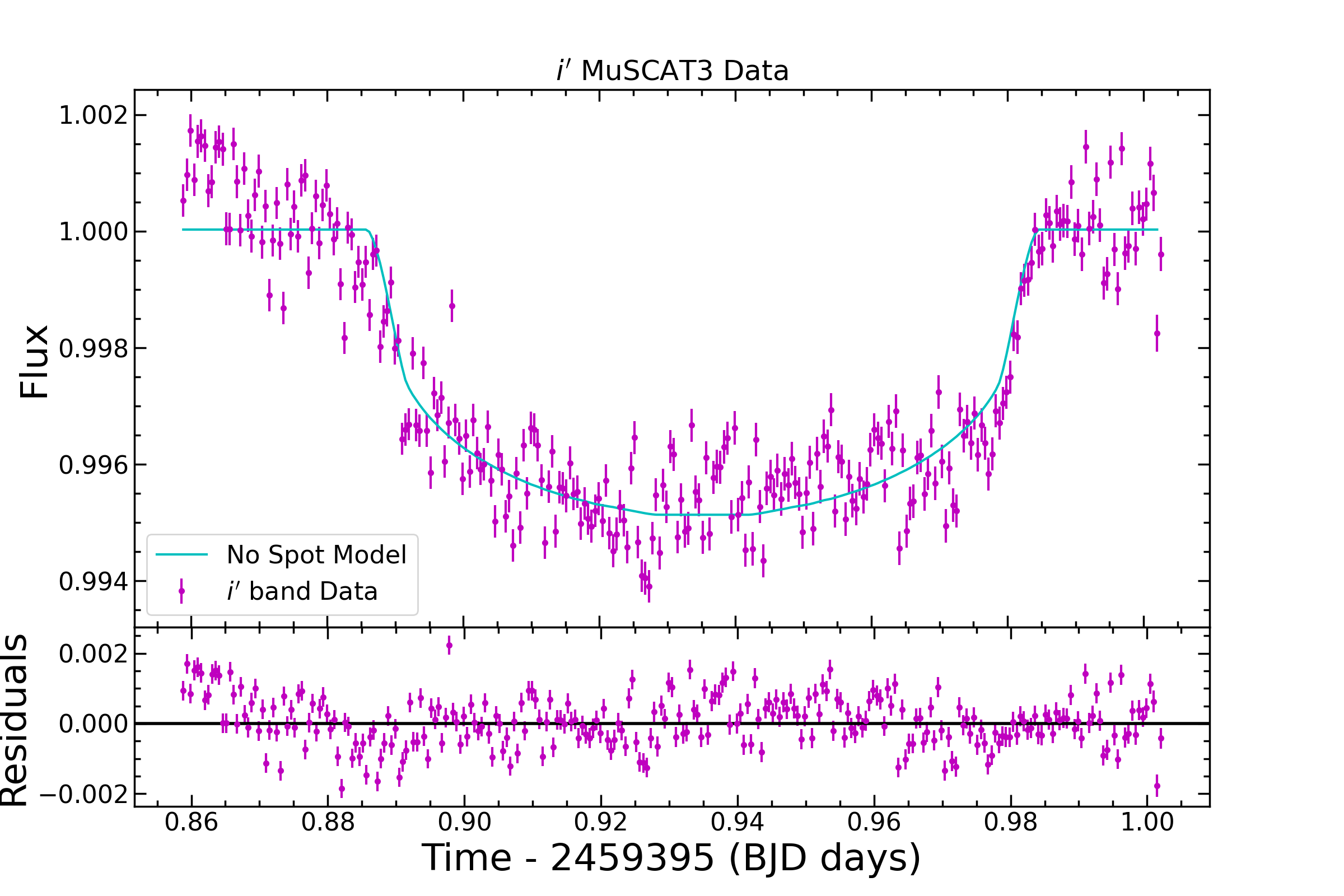}
\includegraphics[width=.4\linewidth]{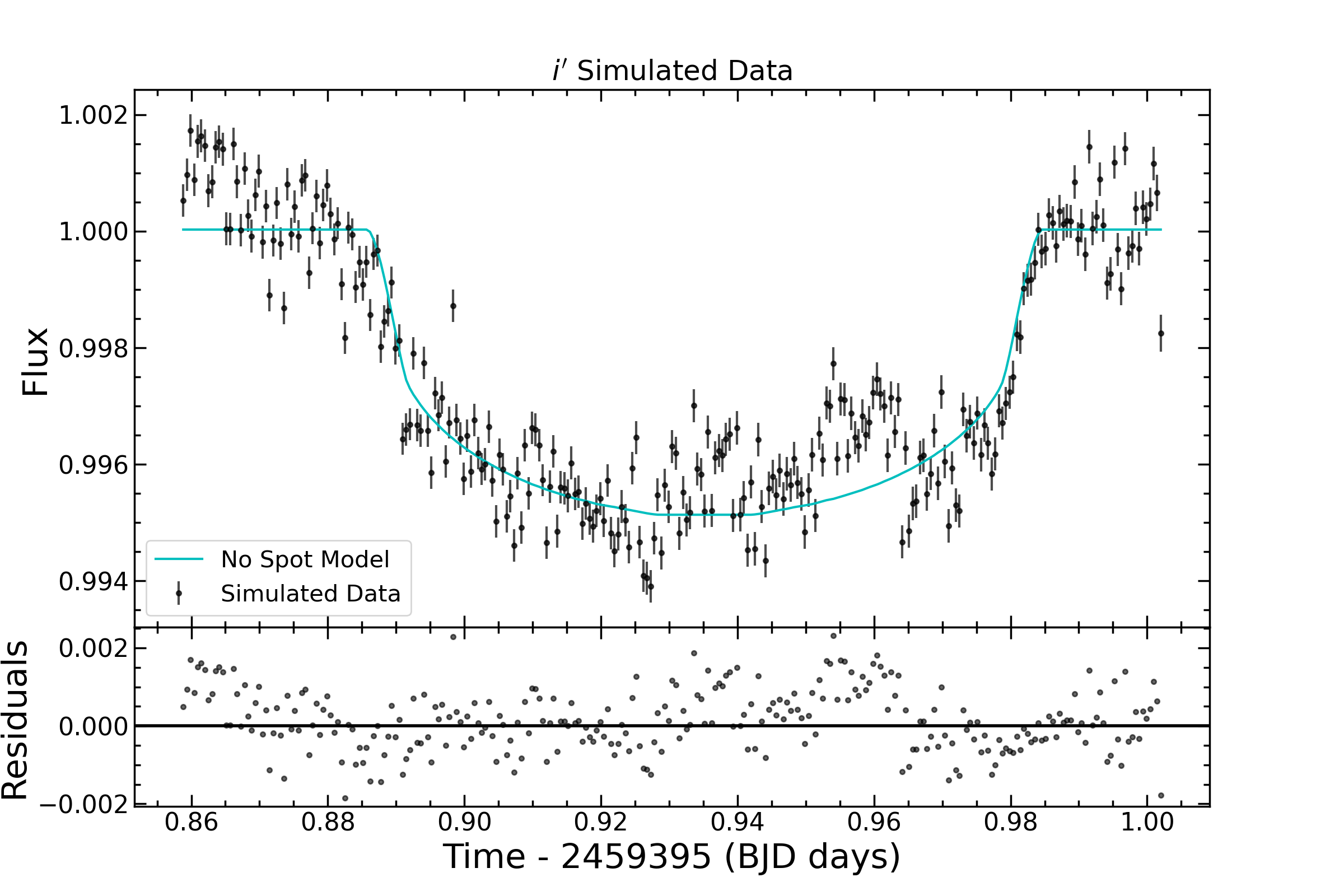}
\includegraphics[width=.4\linewidth]{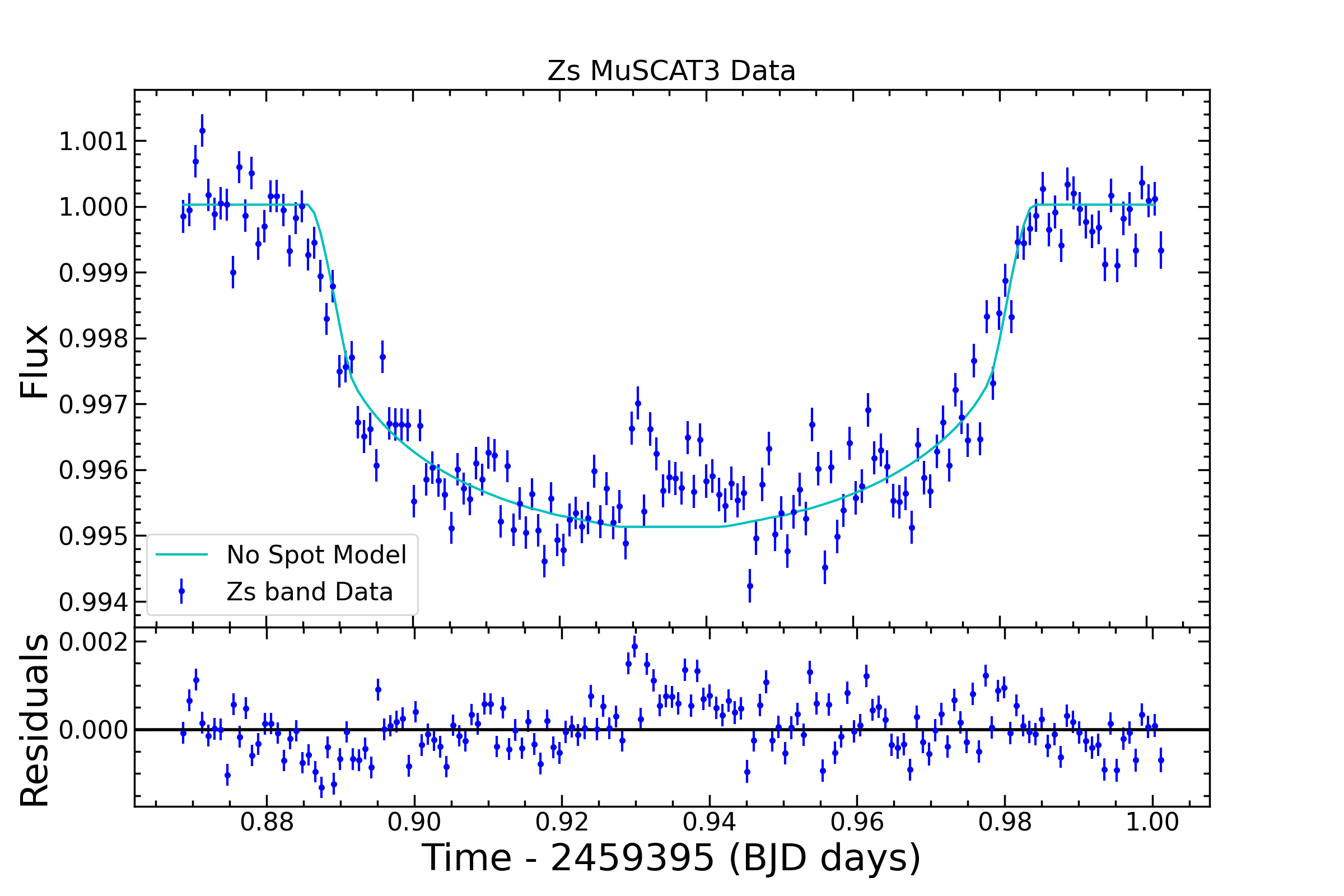}
\includegraphics[width=.4\linewidth]{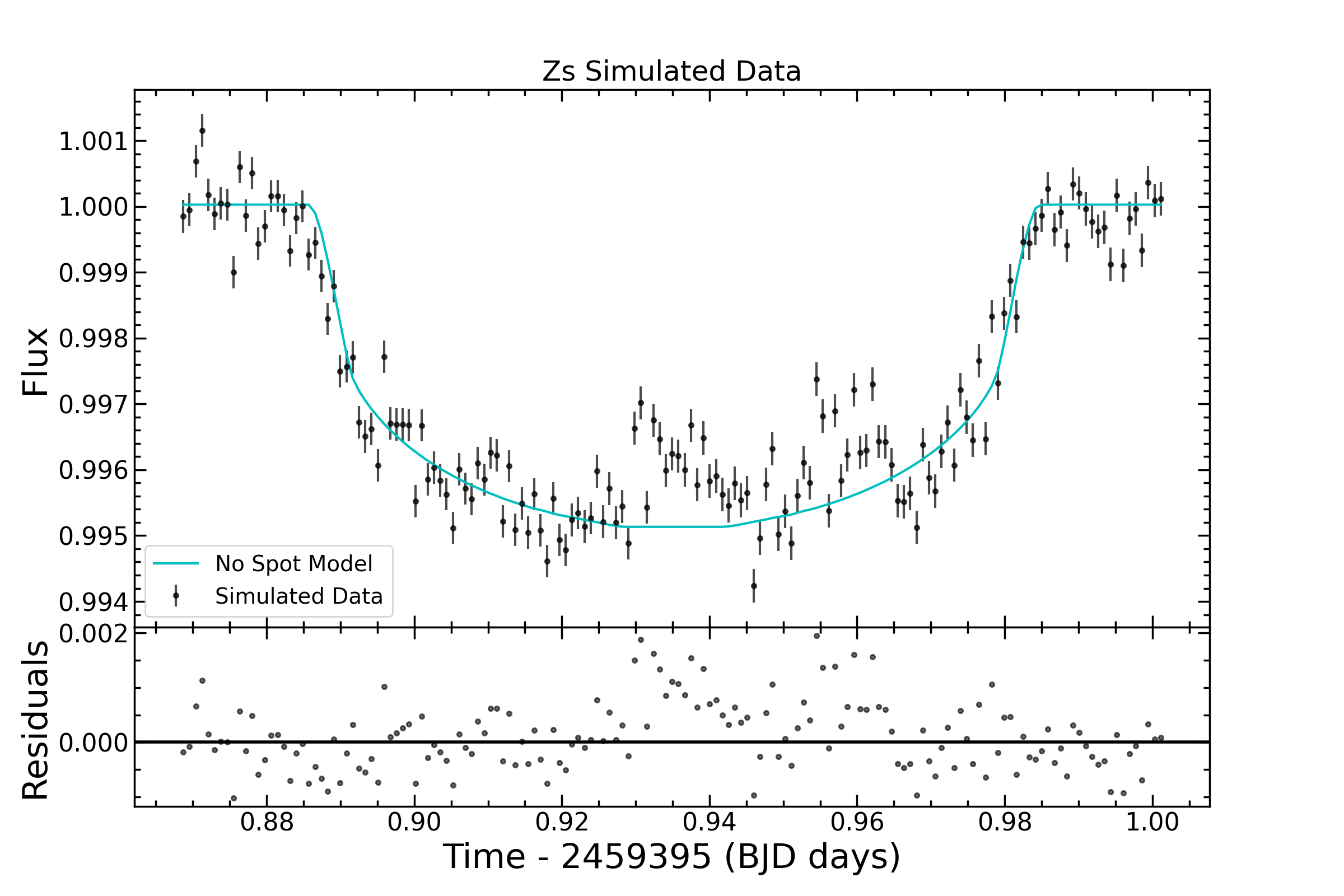}
\caption{{\bf Left:} Data collected by MuSCAT3 instrument for HAT-P-11 b transit obtained on June 29th, 2021. The $g^{\prime}$ band data are shown in the top panel with a precision of 0.58 mmag with an exposure time of 70 s. The $r^{\prime}$ band data are shown below the $g^{\prime}$ with a precision of 0.57 mmag with an exposure time of 40 s. The $i^{\prime}$ band data are shown next with a precision of 0.62 mmag with an exposure time of 40 s. Lastly, the Zs band data are shown in the bottom panel with a precision of 0.54 mmag with an exposure time of 70 s. {\bf Right:} Simulated light curves of LCO data in $g^{\prime}, r^{\prime}, i^{\prime}$ and Zs bands from top to bottom panels respectively assuming the star was spotted as in the \textit{Kepler} light curve in Figure \ref{fig:hat11_transit}. Black points correspond to simulated data with appropriate error bars. The no spot transit model for each filter is shown as a cyan line with the residuals (no spot model - simulated data) shown as black dots in bottom panel of each figure.}
\label{fig:lco_data}
\end{figure*}

Even though we did not observe a starspot crossing during our HAT-P-11 MuSCAT3 transit, we were still able to use the excellent data collected to estimate what we could have observed. In order to simulate what the spots might look like for MuSCAT3, we took the data collected for each filter and subtracted the no spot transit model to get an accurate noise profile for the each filter with the same cadence as the observed data. Note there is a feature in the Zs band data around the transit midpoint that is attributed to noise as this feature is not present in any other filter, which is what would be expected for a starspot crossing event. Then, we added on our corresponding \texttt{STSP} model for a 4500 K starspot for each filter in order to simulate what the spots seen in this transit of HAT-P-11 would look like for the observed noise levels from our MuSCAT3 transit. Our simulated light curves for each filter can be seen on the right hand side of Figure \ref{fig:lco_data} as the grey points with error bars. The real data from the LCO MuSCAT3 transit can be seen in the left hand side of Figure \ref{fig:lco_data} for comparison. As there was a significant bump due to noise in the Zs band around the same location as one of the injected spots, there is a larger than expected first bump in the Zs simulated data due to the real noise of the Zs transit (bottom row of Figure \ref{fig:lco_data}). There is also an portion of the large bump feature in the SDSS $r^{\prime}$ simulated data (second row in Figure \ref{fig:lco_data}) that has higher than expected points due to real noise in the LCO data.

\section{Results and Discussion}\label{sec:res_dis}

\begin{figure*}[t]
%  \begin{subfigure}[t]{.4\textwidth}
\centering
\includegraphics[width=.4\linewidth]{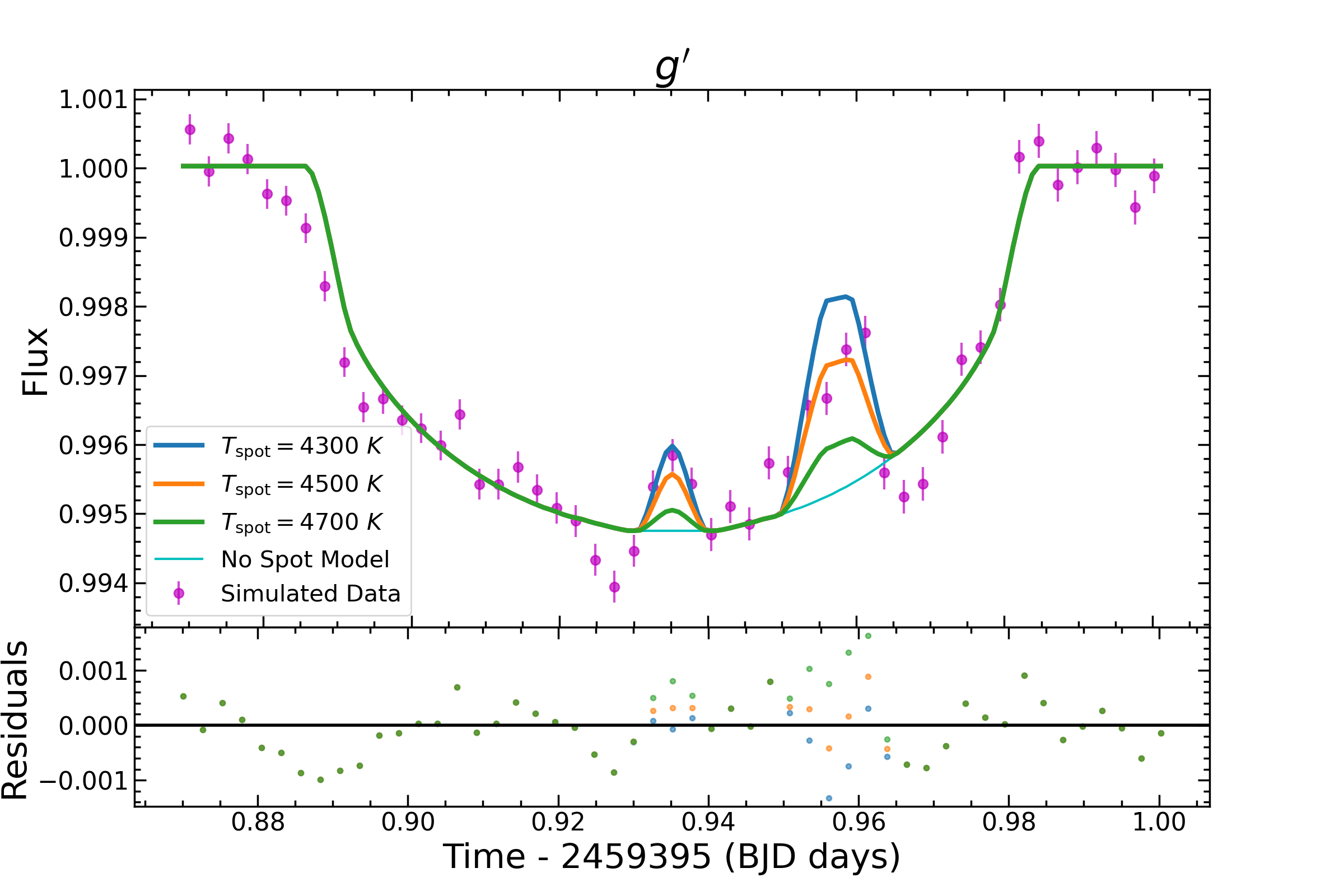}
\includegraphics[width=.4\linewidth]{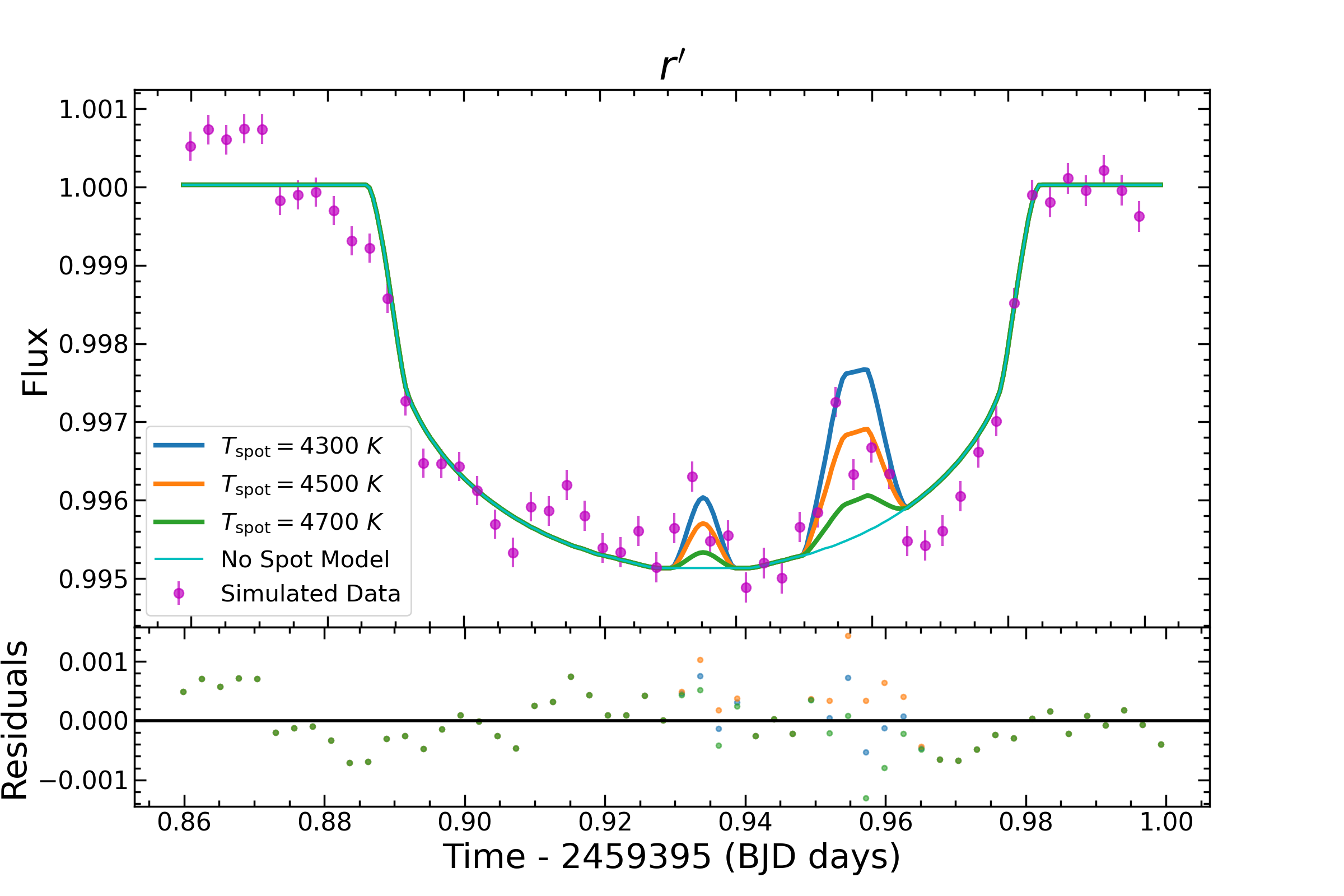}
\includegraphics[width=.4\linewidth]{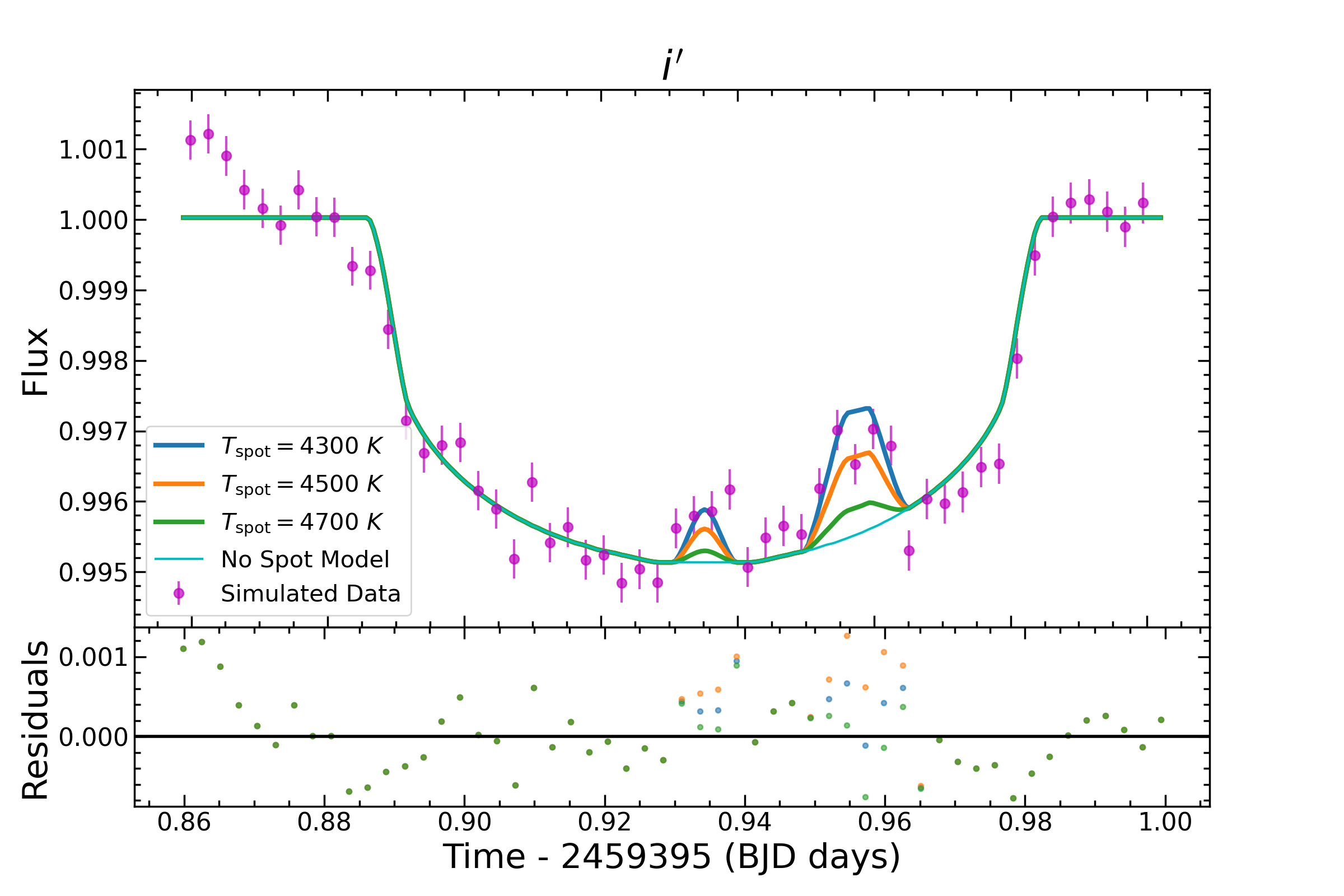}
\includegraphics[width=.4\linewidth]{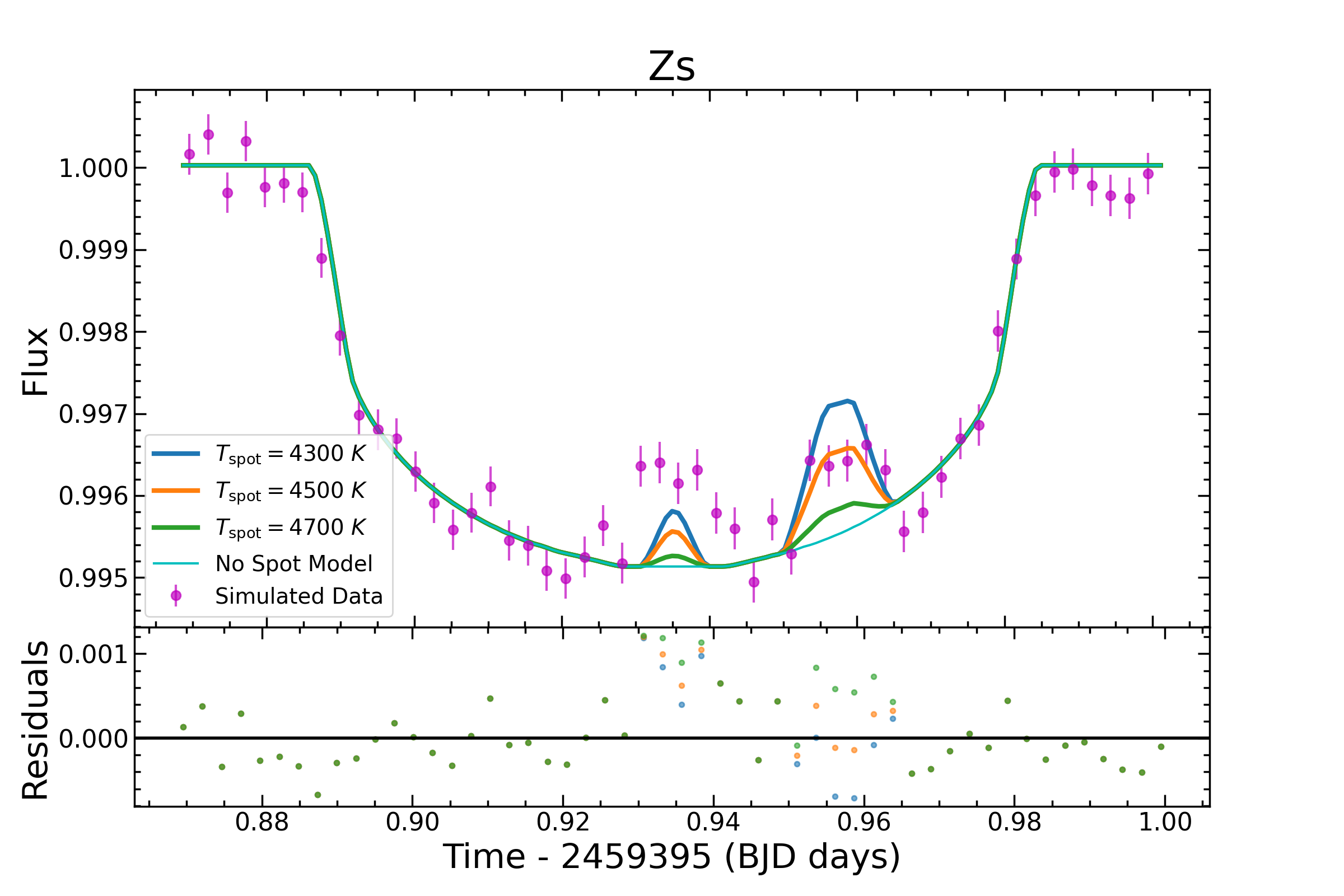}
\caption{Simulated light curves of binned LCO data in $g^{\prime}, r^{\prime}, i^{\prime}$ and Zs bands (top left, top right, bottom left, and bottom right respectively) compared to three \texttt{STSP} models corresponding to $\teff_{spot} =$ 4300 K (blue line), 4500 K (orange line), and 4700 K (green line) with residuals of model - data shown for all three cases in bottom panel. In all cases the $\chi^2$ of the \texttt{STSP} model compared to the simulated data is lowest for $\teff = 4500~K$, though the two closest spot temperature models (4400 and 4600 K) fit within the $\Delta~\chi^2$ of the 4500 K \texttt{STSP} model.}
\label{fig:sim_comp}
\end{figure*}

We will now treat our simulated HAT-P-11 data as the true data we received from MuSCAT3. Thus, we took the grey data points from Figure \ref{fig:lco_data} and modeled them in \texttt{STSP} with the known spot parameters. Keeping the spot parameters the same for all four filters, the only independent variable to change is the contrast for each filter. We modeled eleven spot temperatures for each filter (3700 - 4700 K). Then, we compared all the spot temperature curves to the simulated data and calculated the $\chi^2$ for each temperature. The comparison between the binned simulated data (pink points) and three different spot temperature models (4300 K, 4500 K, and 4700 K) are shown in Figure \ref{fig:sim_comp}. For all four filters, the calculated $\chi^2$ for each spot temperature model compared to the unbinned simulated data is the lowest for the 4500 K model. However, there is some uncertainty in the spot temperature such that the 4400 and 4600 K models have $\chi^2$ values that fall within the $\Delta~\chi^2$ of the 4500 K model. Thus, the determined spot temperature for HAT-P-11's starspots would be 4500 $\pm$ 100 K. 

Since the simulated data that we are modeling was created using the real noise for each filter, it is necessary to discuss each filter individually. For all of the four filters, there is a portion of the observation at the very beginning that is not fit well due to partly cloudy conditions during the observation. For the SDSS $g^{\prime}$ filter (top left panel of Figure \ref{fig:sim_comp}), there is a dip in the real LCO data around 0.925 (2459395 BJD days) that leads to a dip in the binned simulated data (pink points). Additionally, the SDSS $g^{\prime}$ real LCO data has a starspot-like feature around 0.96 (2459395 BJD days) that doesn't appear in any of the other filters and is within the noise level of the transit, so it is likely noise and not a real starspot crossing event. This feature does coincide with the second injected starspot, which creates a larger than expected feature that could match a cooler spot temperature model, like the blue 4300 K spot model in Figure \ref{fig:sim_comp}. Thus, if we only had the SDSS $g^{\prime}$ filter data, a cooler spot temperature might be measured, though the overall lowest $\chi^2$ model is still 4500 K. 

The SDSS $r^{\prime}$ filter (top right panel of Figure \ref{fig:sim_comp}) has one unique aspect in that there was one section of higher noise around 0.955 (2459395 BJD days) in the real LCO noise. This noise again coincides with the start of the second injected starspot causing a higher than expected point in the binned simulated data in Figure \ref{fig:sim_comp} (pink points). However, even with this added noise, the lowest $\chi^2$ spot temperature model is 4500 K as the $\chi^2$ is calculated with respect to the unbinned simulated data compared to the spot temperature models. There is also a lower than expected portion in the simulated data around 0.97 (2459395 BJD days) that is caused by real noise in the LCO data. The SDSS $i^{\prime}$ filter (bottom left panel of Figure \ref{fig:sim_comp}) shows a similar story to the SDSS $r^{\prime}$ band, but for the $i^{\prime}$, the noise is more pronounced starting around 0.93 (2459395 BJD days), which leads to higher than expected bumps in the binned simulated data for both injected starspots. Again, the lowest $\chi^2$ spot temperature model is 4500 K when comparing the models to the unbinned simulated data.

Lastly, the Zs band (bottom right panel of Figure \ref{fig:sim_comp}) has a unique feature around 0.92-0.94 (2459395 BJD days) in the unbinned simulated data. This creates a feature that has a higher amplitude than expected after we injected the smaller starspot into the real LCO noise. This noise is likely real noise (i.e. not a starspot crossing event) that is more pronounced due to this filter being very near-infrared and having the lowest efficiency of all the filters. Because this noise feature is so large, the first smaller feature is only fit with very cool spot temperatures ($>~$3900 K). The second larger injected starspot is fit very well with the 4500 K spot temperature model, and as this feature has less noise in the real LCO data, it is a better feature for comparison. Thus, the best spot temperature model, both by eye and by $\chi^2$, when only considering the second feature is the 4500 K spot model.

 Previously, starspot temperatures have been measured using spectroscopic techniques (see \citet{bergyugina2005} and references therein). This technique has been applied to many different types of stars, and though it has been mostly G and K stars, it has not been used to measure the starspot temperature of HAT-P-11. \citet{morris2019} used a spot temperature difference of $\Delta T_{\rm eff} = 250~K$ in their modeling of spectra obtained with the ARC 3.5-meter telescope at Apache Point Observatory, but that spot temperature was assumed in order to measure the spot covering fraction. \citet{morris2019} indicates that their method would not be able to accurately determine the spot covering fraction (or spot temperature if you assumed the spot covering fraction instead) because their method is not sensitive to spot covering fractions of less than 20\%, and HAT-P-11's maximum spot covering fraction is around 14\% \citep{morris2017}.

\subsection{Limitations of Method}

For this technique to be very successful, there are a few key factors that must be considered. One is that we need high-precision and high-cadence photometry in order to possibly catch a starspot crossing event. Fortunately, we can achieve very high ($<$ 1 mmag) precision from ground based facilities that have diffuser-assisted imaging available. High-precision transit photometry is also available through current and past space-based missions like \textit{Kepler}, K2, TESS, and CHEOPS \citep{cheops}, but this technique does require observations of the same starspot in at least two filters to confirm rather than estimate the spot temperature, which those missions do not have. This leads to the last key factor which is target selection limitations. Since we are currently limited to ground-based high-precision photometry, there are only so many known systems with starspot crossing events that are bright enough to be observed from the ground. However, for those objects, there is no guarantee that a starspot crossing event that can be seen from the ground (i.e. a sufficiently large starspot like the big one in the simulated HAT-P-11 light curves) will occur during the observed transit. Additionally, it may be possible to observe the same starspot region with both TESS and CHEOPS, which would open the target list to many more options.

The limitations of this method were shown quite clearly in this paper, as HAT-P-11 is an ideal target for diffuser-assisted ground based observations due to its brightness. HAT-P-11 also has very well characterized stellar surface features and is known to host starspot crossing events in 95\% of its \textit{Kepler} transits. Additionally, our LCO MuSCAT3 transits had $<$1 mmag precision in every filter. However, most of the starspots on HAT-P-11 are sized more similarly to the smaller bump in Figure \ref{fig:hat11_transit}. As is seen in Figure \ref{fig:sim_comp}, the smaller of the two bumps can only be clearly seen in the binned SDSS $g^{\prime}$ data. Even with the exquisite precision from the LCO data, it would only be possible to catch a large starspot crossing in multiple filters for HAT-P-11 and likely other targets as well. 

\subsection{Future Work}

With the JWST \citep{jwst} and upcoming Pandora SmallSat \citep{quintana2021} missions, there will soon be an influx of transmission spectroscopy done on exoplanet atmospheres around a wide range of host stars. Thus, it is becoming increasingly more important to be able to characterize and model stellar surface features on stars with a wide range of temperatures and rotation periods. In the  modeling of starspots and faculae, the temperature of the active region is necessary in order to accurately determine the size of the region. With our introduced method of theoretically determining the contrast of the starspots for a range of spot temperatures, it is possible to determine within 100 K the temperature of the spots with only one transit in at least two filters, assuming there is a starspot crossing event observed during the transit. If there is a persistent starspot that is observed in multiple transits and with different filters, then this method could also be successful in determining the spot temperature, though any observational difference (e.g. weather or seeing shifts during the transit) could influence the observed spot occultation as well (see \citet{3884} for example). In either case, this method is not observationally expensive as only one or two transits for each object is required. 

Even with only single band photometry, it would be possible if there are simultaneous spectra taken of the object to measure the spot temperature. This could be done using a combination of modeling the spots using the photometry, either with in-transit spot occultations or out-of-transit photometric modeling (see \citet{wisniewski2019} for reference) which provide a simultaneous measurement of the spot covering fraction. With this simultaneous spot covering fraction measurement, then the concurrent spectra can be modelled using a two temperature spectral decomposition framework. This type of light curve modeling combined with spectral decomposition modeling has been done by \citet{gosnell2022} on a subgiant star with a large covering fraction of 32\%, although their work was done for non-simultaneous photometric and spectroscopic data. However, with a mission like the upcoming Pandora SmallSat mission \citep{quintana2021}, it will be possible to perform this type of method with simultaneous single-band photometry and near-IR spectra.

\section{Conclusions}\label{sec:conc}

Starspot properties, such as their temperature, are important components to understanding both stellar magnetic dynamo theory and exoplanetary transmission spectroscopy. Historically, starspot temperatures have been measured using spectroscopic techniques that leverage different molecular bands that appear only at certain temperatures, but these methods work best for G and K stars with high starspot covering fractions (see \citet{bergyugina2005,morris2019} and references therein). For HAT-P-11 which has a maximum covering fraction of 14\% or for lower temperature stars like M dwarfs, these spectroscopic methods are not ideal. Thus, we have instead leveraged a starspot crossing during a transit to devise a method to measure the spot temperature using high-precision photometry. Using high precision, multi-filter photometry, we have demonstrated the ability to determine the spot temperature to within 100 K if there is a starspot occultation event using a HAT-P-11b transit obtained using the MuSCAT3 instrument on LCO's 2.0-meter FTN telescope. This method can be used for any two filters with different enough contrasts, though SDSS $g^{\prime}$ and SDSS $i^{\prime}$ created the largest signal difference with the highest cadence in our work and for our object. Future missions such as Pandora will provide simultaneous photometric and spectroscopic data during transiting events, which will allow for even more measurements of spot temperatures and covering fractions. 

\section{Acknowledgements}

We acknowledge support from NSF grant AST-1907622. This paper is based on observations made with the MuSCAT3 instrument, developed by the Astrobiology Center and under financial supports by JSPS KAKENHI (JP18H05439) and JST PRESTO (JPMJPR1775), at Faulkes Telescope North on Maui, HI, operated by the Las Cumbres Observatory. This research has made use of the NASA Exoplanet Archive, which is operated by the California Institute of Technology, under contract with the National Aeronautics and Space Administration under the Exoplanet Exploration Program. This paper includes data collected by the \textit{Kepler} mission and obtained from the MAST data archive at the Space Telescope Science Institute (STScI). Funding for the \textit{Kepler} mission is provided by the NASA Science Mission Directorate. STScI is operated by the Association of Universities for Research in Astronomy, Inc., under NASA contract NAS 5–26555. This research has made use of the Spanish Virtual Observatory (https://svo.cab.inta-csic.es) project funded by MCIN/AEI/10.13039/501100011033/ through grant PID2020-112949GB-I00.  

GS acknowledges support provided by NASA through the NASA Hubble Fellowship grant HST-HF2-51519.001-A awarded by the Space Telescope Science Institute, which is operated by the Association of Universities for Research in Astronomy, Inc., for NASA, under contract NAS5-26555.

CIC acknowledges support by NASA Headquarters through an appointment to the NASA Postdoctoral Program at the Goddard Space Flight Center, administered by USRA through a contract with NASA.

\newpage
\bibliography{references}

\section{Appendix}\label{sec:appendix}
%Should we just drop this? I already dropped it from the text and I don't know that it's necessary 
% \subsection{Kepler-63 Data}

% We observed a full transit of Kepler-63b on May 25th, 2021 with MuSCAT3. Kepler-63 is a well-studied, young solar analogue with a hot gas giant planet that exhibits large amounts of starspot activity \citep{netto2020}. Kepler-63 has a SDSS $g^{\prime}$ magnitude of 12, which makes the exposure times longer than for HAT-P-11 by a factor of around 1.5 with the aid of the available diffusers. Unfortunately, the weather on this night was partly cloudy for the length of the transit with one very cloudy section close to the middle of the transit (see Figure \ref{fig:kep63_bin}). Thus, the data quality is too poor to determine if a starspot crossing occurred on this night. Additionally, due to the poor weather, we felt the data were not sufficiently clean enough to perform the same procedure as done above for HAT-P-11. 

% \begin{figure}[h]
% \centering
% \includegraphics[width=0.9\textwidth]{figures/K63_contrast_all_bin.png}
% \caption{Data from LCO MuSCAT3 from 05-25-2021 for Kepler-63 binned to 200 seconds for SDSS $g^{\prime}$ (green), 120 seconds for SDSS $r^{\prime}$ (red), 240 seconds for SDSS $i^{\prime}$ (magenta) and 340 seconds for Zs (blue) with all filters plotted on top of each other. The bottom panel shows the residuals of the no spot model (cyan line) minus the binned data points. }
% \label{fig:kep63_bin}
% \end{figure}

\subsection{Polynomial Fits to Contrast Curves}
\begin{figure}[h]
\centering
\includegraphics[width=0.9\textwidth]{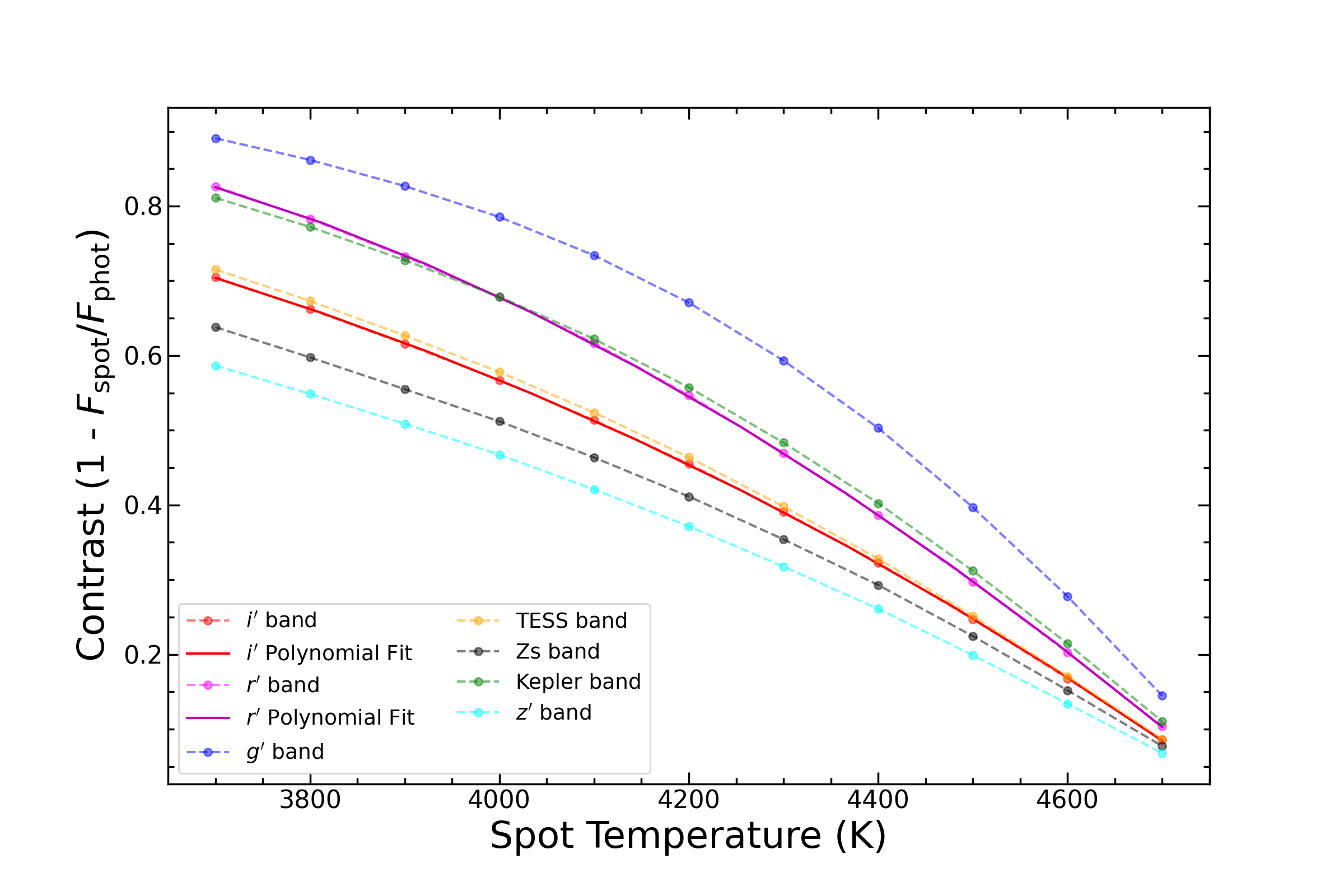}
\caption{Data for calculated HAT-P-11 contrast versus spot temperature for the following filters are shown as data points with dashed lines: SDSS $g^{\prime}$ (blue), SDSS $r^{\prime}$ (magenta), SDSS $i^{\prime}$ (red), SDSS $z^{\prime}$ (cyan), TESS (orange), \textit{Kepler} (green), and OAO Zs (black). Two polynomial fits are shown for the SDSS $r^{\prime}$ and SDSS $i^{\prime}$ as solid colored lines. }
\label{fig:hat11_poly}
\end{figure}

In addition to calculating the contrast for every spot temperature in every filter as described in Section \ref{sec:data}, we have fit a polynomial to the contrast values shown in Figure \ref{fig:con_curves}. For our polynomial fitting, we are interested in the ability to input any spot temperature and calculate a contrast for the chosen filter for HAT-P-11. Thus, we fit the polynomial to our calculated contrast data versus spot temperature rather than wavelength as shown in Figure \ref{fig:hat11_poly}. We have shown the contrast versus spot temperature for the following filters as data points and dashed lines: SDSS $g^{\prime}$ (blue), SDSS $r^{\prime}$ (magenta), SDSS $i^{\prime}$ (red), SDSS $z^{\prime}$ (cyan), TESS (orange), \textit{Kepler} (green), and OAO Zs (black) with the example polynomial fits shown for the SDSS $r^{\prime}$ and SDSS $i^{\prime}$ as solid colored lines. All the polynomial fits have been done using third order polynomials of the form given in Equation (2) where $x$ stands for spot temperature, $[a, b, c, d]$ are the polynomial coefficients, and $p(x)$ provides the contrast for that spot temperature. The polynomial coefficients for each filter are given in Table \ref{tab:poly}. These polynomial curves will thus allow for the ability to choose any spot temperature ($x$) in any of the listed filters and calculate the contrast for a star with the same effective photosphere temperature and surface gravity as HAT-P-11 (\teff = 4800 K and \logg = 4.5).

\begin{equation}
    p(x) = ax^3 + bx^2 + cx +d
\end{equation}

\begin{center}
\begin{table}[]
\centering
\caption{Polynomial Coefficients}
\begin{tabular}{ccccc}
\hline
Filter & a & b & c & d  \\
\hline
SDSS $g^{\prime}$ & 1.846e-10 & 1.715e-06 & -5.339e-03 & 6.514 \\
SDSS $r^{\prime}$ & 2.932e-11 & -6.964e-07 & 3.568e-03 & -4.329 \\
SDSS $i^{\prime}$ & -2.979e-11 & 1.357e-07 & -1.750e-04 & 1.003 \\
SDSS $z^{\prime}$ & -1.426e-11 & 1.708e-09 & 2.249e-04 & 4.526e-01\\
Zs & -5.947e-11 & 5.331e-07 & -1.878e-03 & 3.297 \\
TESS & -1.414e-11 & -7.307e-08 & 7.371e-04 & -2.964e-01 \\
\textit{Kepler} & -4.579e-11 & 1.909e-07 & 1.298e-04 & 3.452e-02 \\\hline
\end{tabular} 
\label{tab:poly}
\end{table}
\end{center}

\end{document}